\documentclass[%
 reprint,
 amsmath,amssymb,amsthm,
 nofootinbib,
 aps,
]{revtex4-1}

\usepackage{dcolumn}
\usepackage{bm}
\usepackage[mathlines]{lineno}
\usepackage{graphicx}
\usepackage{csquotes}

\newtheorem{theorem}{Theorem}

\begin{document}

\title{Quantum Mechanics from Relational Properties \\Part II: Measurement and EPR}

\author{Jianhao M. Yang}
\email{jianhao.yang@alumni.utoronto.ca}
\affiliation{
San Diego, CA 92121, USA
}

\date{\today}

\begin{abstract}
Quantum measurement and quantum operation theory is developed here by taking the relational properties among quantum systems, instead of the independent properties of a quantum system, as the most fundamental elements. By studying how the relational probability amplitude matrix is transformed and how mutual information is exchanged during measurement, we derive the formulation that is mathematically equivalent to the traditional quantum measurement theory. More importantly, the formulation results in significant conceptual consequences. We show that for a given quantum system, it is possible to describe its time evolution without explicitly calling out a reference system. However, description of a quantum measurement must be explicitly relative. Traditional quantum mechanics assumes a super observer who can instantaneously know the measurement results from any location. For a composite system consists space-like separated subsystems, the assumption of super observer must be abandoned and the relational formulation of quantum measurement becomes necessary. This is confirmed in the resolution of EPR paradox. Information exchange is relative to a local observer in quantum mechanics. Different local observers can achieve consistent descriptions of a quantum system if they are synchronized on the information regarding outcomes from any measurement performed on the system. It is suggested that the synchronization of measurement results from different observers is a necessary step when combining quantum mechanics with the relativity theory.
\begin{description}
\item[PACS numbers] 03.65.Ta, 03.65.-w
\item[Keywords] Relational Quantum Mechanics, Quantum Measurement, Entanglement, EPR
\end{description}
\end{abstract}
\maketitle

\section{Introduction}
\label{intro}
Quantum mechanics was originally developed as a physical theory to explain the experimental observations of a quantum system in a measurement. In the early days of quantum mechanics, Bohr had emphasized that the description of a quantum system depends on the measuring apparatus~\cite{Bohr, Bohr35, Jammer74}. In more recent development of quantum interpretations, the dependency of a quantum state on a reference system was further recognized. The relative state formulation of quantum mechanics~\cite{Everett, Wheeler57, DeWitt70} asserts that a quantum state of a subsystem is only meaningful relative to a given state of the rest of the system. Similarly, in developing the theory of decoherence induced by environment~\cite{Zurek82, Zurek03, Schlosshauer04}, it is concluded that correlation information between two quantum systems is more basic than the properties of the quantum systems themselves. Relational Quantum Mechanics (RQM) further suggests that a quantum system should be described relative to another system, there is no absolute state for a quantum system~\cite{Rovelli96, Rovelli07}. Quantum theory does not describe the independent properties of a quantum system. Instead, it describes the relation among quantum systems, and how correlation is established through physical interaction during measurement. The reality of a quantum system is only meaningful in the context of measurement by another system.

The idea of RQM is thought provoking. It essentially implies two aspects of relativity. First, since a quantum system should be described relative to another system, the RQM theory claims that the relational properties between two quantum systems are more basic than the independent properties of one system. The relational properties instead of the independent properties of a quantum system should be considered as a starting point for constructing the formulation of quantum mechanics. Questions associated with this aspect of relativity include how to quantify the relational properties between the observed system and the reference system? How to reconstruct a quantum mechanics theory from relational properties?  Second, the relational properties themselves are relative to a reference system. The reference system is arbitrarily selected. It can be an apparatus in a measurement setup, or another system in the environment. A quantum system can be described differently relative to different reference systems. The reference system itself is also a quantum system, which is called a quantum reference frame (QRF). There are extensive research activities on QRF, particularly how to ensure consistent descriptions when switching QRFs~\cite{QRF1, QRF2, QRF3, QRF4, QRF5, QRF6, QRF7, QRF8, QRF9, QRF10, QRF11, QRF12, QRF13, QRF14, QRF15, QRF16, Brukner, Hoehn2018}. Even if the reference system is selected and the relational properties are quantified, there can be multiple observers. It is possible that another observer does not have complete information of the interaction (or, measurement) results between observed system and the reference system. In such case she can describe the observed system differently using a different set of relational properties between observed system and the observing system. It is in this sense that we say the relational properties themselves are observer-dependent. This is indeed the main thesis of Ref.~\cite{Rovelli96}. There are several fundamental questions associated with the second aspect of relativity: Is such description compatible with traditional quantum theory which appears as observer independent (at least the non-relativistic quantum mechanics)? If a quantum system should be described as observer-dependent, how is the objectivity of a physical reality preserved? In what condition RQM can provide results that are different from traditional quantum mechanics? 

Recently we proposed a formulation to address the first aspect of RQM~\cite{Yang2017}. In this formulation, the relational properties between the two quantum systems are the most fundamental elements to formulate quantum mechanics. In addition, a novel framework of calculating the probability of an outcome when measuring a quantum system is proposed by modeling the bidirectional probe-response interaction process in a measurement. The fundamental relational property is defined as relational probability amplitude. The probability of a measurement outcome is proportional to the summation of probability amplitude product from all alternative measurement configurations. The properties of quantum systems, such as superposition and entanglement, are manifested through the rules of counting the alternatives. As a result, the framework gives mathematically equivalent formulation to Born's rule. Wave function is found to be summation of relational probability amplitudes, and Schr\"{o}dinger Equation is derived when there is no entanglement in the relational probability amplitude matrix. Although the relational probability amplitude is the most basic properties, there are mathematical tools such as wave function and reduced density matrix that describe the observed system without explicitly called out the measuring system. Thus, the formulation in Ref.~\cite{Yang2017} is mathematically compatible to the traditional quantum mechanics. 

This paper has two goals. The first goal is to extend the formulation presented in Ref.~\cite{Yang2017} to the quantum measurement theory. One of the key concept in this formulation is the entanglement measure of the relational probability amplitude matrix. The entanglement measure quantifies the difference between time evolution and measurement. When there is change in the entanglement entropy, we expect to derive the quantum measurement theory, which is missing in Ref.~\cite{Yang2017}. This paper intends to complete the formulation for quantum measurement and quantum operation in the context of RQM. The reformulation is mathematically equivalent to the traditional quantum measurement theory and open quantum system theory. The second goal of this paper is to investigate how the measurement theory is perceived from different observers. Here we assume these different observers use the same reference frame. How the measurement theory is transformed when switching QRF is not in the scope of this paper. Therefore, this investigation only partially addresses the second aspect of RQM. But as shown later, it already results in several important conceptual implications. For instance, We assert that for a given quantum system, description of its time evolution can be implicitly relative, while description of a quantum operation must be explicitly relative. Information exchange is relative to a local observer in quantum mechanics. The assumption of Super Observer should be abandoned, so as the notion of observer independent description of physical reality. Different local observers can achieve consistent descriptions of a quantum system if they are synchronized on the outcomes from any measurement performed on the system. Traditional quantum mechanics was originally developed to explain observation results from microscopic system that is much smaller than the measuring apparatus. In those situations, RQM and traditional quantum mechanics are practically equivalent. However, for a composite system that is spatially much larger than a typical apparatus, the necessity of RQM formulation become obvious. This is clearly illustrated in the analysis of EPR paradox~\cite{EPR}. The paradox is seemingly inevitable in traditional quantum mechanics but can be resolved by removing the assumption of the Super Observer who knows measurement results instantaneously from local observer from any location. It is suggested that the synchronization of measurement results from different observers is a necessary step when combining quantum mechanics with Relativity Theory.

The works presented here is inspired by the main idea of the original RQM theory~\cite{Rovelli96}. However, there are significant advancements compared with the original RQM theory. They are summarized in Section \ref{sec:conclusion}.

The paper is organized as following. Firstly we briefly review the relational formulation of quantum mechanics in Section \ref{sec:RQM}. In Section \ref{sec:measurement} we present the measurement theory based on the relational formulation and in Section \ref{sec:generalOp} the formulation is extended to general quantum operation. It turns out that Schr\"{o}dinger Equation, formulations for selective and non-selective measurement, can all be derived from the general quantum operation. Section \ref{criteria} analyzes the criteria on whether a quantum process must be described by calling out the observer explicitly. The result is applied to resolve the EPR paradox in Section \ref{subsec:EPR}. Lastly, the conceptual consequences, the potential applications of this formulation, and the conclusion remarks are presented in Section \ref{sec:conclusion}. 

\section{Relational Formulation of Quantum Mechanics}
\label{sec:RQM}

\subsection{Terminologies}
\label{subsec:definition}
A \textit{Quantum System}, denoted by symbol $S$, is an object under study and follows the quantum mechanics postulates~\cite{Yang2017}. An \textit{Apparatus}, denoted as $A$, can refer to the measuring devices, the environment that $S$ is interacting with, or the system from which $S$ is created. All systems are quantum systems, including any apparatus. Depending on the selection of observer, the boundary between a system and an apparatus can change. For example, in a measurement setup, the measuring system is an apparatus $A$, the measured system is $S$. However, the composite system $S+A$ as a whole can be considered a single system, relative to another apparatus $A'$. In an ideal measurement to measure an observable of $S$, the apparatus is designed in such a way that at the end of the measurement, the pointer state of $A$ has a distinguishable, one to one correlation with the eigenvalue of the observable of $S$.

The definition of \textit{Observer} is associated with an apparatus. An observer, denoted as $\cal{O}$, is an intelligence who can operate and read the pointer variable of the apparatus. Whether or not this observer is a quantum system is irrelevant in our formulation. However, there is a restriction that is imposed by the principle of locality. An observer is defined to be physically local to the apparatus he associates with. This prevents the situation that $\cal{O}$ can instantaneously read the pointer variable of the apparatus that is space-like separated from $\cal{O}$. Receiving the information from $A$ at a speed faster than the speed of light is prohibited. An observer cannot be associated with two or more apparatuses in the same time if these apparatuses are space-like separated. 

In the traditional quantum measurement theory proposed by von Neumann~\cite{Neumann}, both the quantum system and the measuring apparatus follow the same quantum mechanics laws. Von Neumann further distinguished two separated measurement stages, Process 1 and Process 2. Mathematically, an ideal measurement process is expressed as
\begin{equation}
    \label{measurement}
    \begin{split}
    |\Psi\rangle_{SA} & = |\psi_S\rangle|a_0\rangle = \sum_i c_i|s_i\rangle|a_0\rangle \\
    & \longrightarrow \sum_i c_i|s_i\rangle|a_i\rangle \longrightarrow |s_n\rangle|a_n\rangle
    \end{split}
\end{equation}
Initially, both $S$ and $A$ are in a product state described by $|\Psi\rangle_{SA}$. In Process 2, referring to the first arrow in Eq.(\ref{measurement}), the quantum system $S$ and the apparatus $A$ interact. However, as a combined system they are isolated from interaction with any other system. Therefore, the dynamics of the total system is determined by the Schr\"{o}dinger Equation. Process 2 establishes a one to one correlation between the eigenstate of observable of $S$ and the pointer state of $A$. After Process 2, there are many possible outcomes to choose from. In the next step which is called Process 1, referring to the second arrow in Eq.(\ref{measurement}), one of these possible outcomes (labeled with eigenvalue $n$) emerges out from the rest\footnote{Traditional quantum mechanics does not provide a theoretical description of Process 1. In the Copenhagen Interpretation, this is considered as the ``collapse' of the wave function into an eigenstate of the measured observable. The nature of this wave function collapse has been debated over many decades. Recent interpretations of quantum theory advocate that the wave function simply encodes the information that an observe can describe on the quantum system. Therefore, it is an epistemic, rather than ontological, variable. In this view, the collapse of wave function is just an update of the observer's description on the condition of knowing the measurement outcome. For example, Quantum Bayesian theory~\cite{Fuchs02, Fuchs13} formulates how Bayesian theorem can be utilized to describe such process. The relational argument of the wave function ``collapse" is presented in Section \ref{sec:measurement}.}. An observer knows the outcome of the measurement via the pointer variable of the apparatus. Both systems encode information each other, allowing an observer to infer measurement results of $S$ by reading pointer variable of $A$. The key insight learned here is that quantum measurement is a question-and-answer bidirectional process. The measuring system interacts (or, disturbs) the measured system. The interaction in turn alters the state of the measuring system. As a result, a correlation is established, allowing the measurement result for $S$ to be inferred from the pointer variable of $A$. 

A \textit{Quantum Reference Frame (QRF)} is a quantum system where all the descriptions of the relational properties between $S$ and $A$ is referred to. There can be multiple QRFs. How the descriptions are transformed when switching QRFs is not in the scope of this study. But we expect the theories developed in Ref.~\cite{Brukner, Hoehn2018, Yang2020} can be applicable here. In this paper, we only consider the description relative to one QRF, denoted as $F$. It is also possible to choose $A$ as the reference frame. In that case, $F$ and $A$ are the same quantum system in a measurement~\cite{Brukner, Yang2020}. Fig. 1 shows a schematic illustration of the entities in the relational formulation.

\begin{figure}
\begin{center}
\includegraphics[scale=2.2]{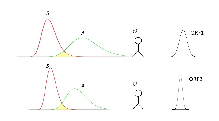}
\caption{Schematic illustration of the entities for the terminologies. The overlapping of the measured system $S$ and apparatus $A$ represents there is interaction in a measurement. The relational properties between $S$ and $A$ must be described relative a QRF. There can be multiple QRFs. Selecting a different QRF, $\cal{O}$ can have a different description of the relational properties in a quantum measurement event.}
\label{fig:1}       
\end{center}
\end{figure}

A \textit{Quantum State} of $S$ describes the complete information an observer $\cal{O}$ can know about $S$. A quantum state encodes the relational information of $S$ relative to $A$ or other systems that $S$ previously interacted with~\cite{Rovelli07}. The information encoded in the quantum state is the complete knowledge an observer can say about $S$, as it determines the possible outcomes of next measurement. As we will explain later, the state for a quantum system is not a fundamental concept. Instead, it is a derivative concept from the relational properties.

\subsection{Basic Formulation}
This section briefly describes the framework of relational formulation of quantum mechanics~\cite{Yang2017}. the framework is based on a detailed analysis of the interaction process during measurement of a quantum system. First, from experimental observations, a measurement of a variable on a quantum system yields multiple possible outcomes randomly. Each potential outcome is obtained with a certain probability. We call each measurement with a distinct outcome a quantum event. Denote these alternatives events with a set of kets $\{|s_i\rangle\}$ for $S$, where ($i=0,\ldots ,N-1)$, and a set of kets $\{|a_j\rangle\}$ for $A$, where ($j=0,\ldots ,M-1)$. A potential measurement outcome is represented by a pair of kets $(|s_i\rangle, |a_j\rangle)$. Second, a physical measurement is a bidirectional process, the measuring system and the measured system interact and modify the state of each other.  The probability of finding a potential measurement outcome represented by a pair of kets $(|s_i\rangle, |a_j\rangle)$, $p_{ij}$, should be calculated by modeling such bidirectional process. This implies $p_{ij}$ can be expressed as product of two numbers,
\begin{equation}
\label{probprod}
    p_{ij} \propto Q_{ji}^{AS}R_{ij}^{SA}. 
\end{equation}
$Q_{ji}^{AS}$ and $R_{ij}^{SA}$ are not necessarily real non-negative number since each number alone only models a unidirectional process which is not a complete measurement process. On the other hand, $p_{ij}$ is a real non-negative number since it models an actual measurement process. To satisfy such requirement, we further assume
\begin{equation}
\label{conjugate}
    Q_{ji}^{AS} = (R_{ij}^{SA})^*.
\end{equation}
Written in a different format, $Q_{ji}^{AS} = (R^{SA})^\dag_{ji}$. This means $Q^{AS} = (R^{SA})^\dag$. Eq.(\ref{probprod}) then becomes 
\begin{equation}
    p_{ij} = |R^{SA}_{ij}|^2/\Omega
\end{equation}
where $\Omega$ is a real number normalization factor. $Q_{ji}^{AS}$ and $R_{ij}^{SA}$ are called \textit{relational probability amplitudes}. Given the relation in Eq.(\ref{conjugate}), we will not distinguish the notation $R$ versus $Q$, and only use $R$. The relational matrix $R^{SA}$ gives the complete description of $S$. It provides a framework to derive the probability of future measurement outcome.  

$R_{ij}^{SA}$ can be explicitly calculated using the path Integral formulation~\cite{Yang2021}. In the context of path integral,  $R^{SA}_{ij}$ is defined as the sum of quantity $e^{iS_p/\hbar}$, where $S_p$ is the action of the composite system $S+A$ along a path. Physical interaction between $S$ and $A$ may cause change of $S_p$, which is the phase of the probability amplitude. But $e^{iS_p/\hbar}$ itself is a probabilistic quantity. Although $R^{SA}_{ij}$ is a probability amplitude, not a probability real number, we assume it follows certain rules in the classical probability theory, such as multiplication rule, and sum of alternatives in the intermediate steps.

The set of kets $\{|s_i\rangle\}$, representing distinct measurement events for $S$, can be considered as \textit{eigenbasis} of Hilbert space ${\cal H}_S$ with dimension $N$, and $|s_i\rangle$ is an eigenvector. Since each measurement outcome is distinguishable, $\langle s_i|s_j\rangle = \delta_{ij}$. Similarly, the set of kets $\{|a_j\rangle\}$ is eigenbasis of Hilbert space ${\cal H}_A$ with dimension $N$ for the apparatus system $A$. The bidirectional process $|a_j\rangle \rightleftharpoons |s_i\rangle$ is called a \textit{potential measurement configuration}. A potential measurement configuration comprises possible eigen-vectors of $S$ and $A$ that involve in the measurement event, and the relational weight quantities. It can be represented by $\Gamma_{ij}: \{|s_i\rangle, |a_j\rangle, R^{SA}_{ij}, Q^{AS}_{ji}\}$.

To derive the properties of $S$ based on the relational $R$, we examine how the probability of measuring $S$ with a particular outcome of variable q is calculated. It turns out such probability is proportional to the sum of weights from all applicable measurement configurations, where the weight is defined as the product of two relational probability amplitudes corresponding to the applicable measurement configuration. Identifying the applicable measurement configuration manifests the properties of a quantum system. For instance, before measurement is actually performed, we do not know that
which event will occur to the quantum system since it
is completely probabilistic. It is legitimate to generalize the potential measurement configuration as $|a_j\rangle \rightarrow |s_i\rangle \rightarrow |a_k\rangle$. In other words, the measurement configuration in the joint Hilbert space starts from $|a_j\rangle$, but can end at $|a_j\rangle$, or any other event, $|a_k\rangle$. Indeed, the most general form of measurement configuration in a bipartite system can be $|a_j\rangle \rightarrow |s_m\rangle \rightarrow |s_n\rangle \rightarrow |a_k\rangle$. Correspondingly, we generalize Eq.(\ref{probprod}) by introducing a quantity for such configuration,
\begin{equation}
\label{micAction}
W_{jmnk}^{ASSA} = Q^{AS}_{jm}R^{SA}_{nk} = (R^{SA}_{mj})^*R^{SA}_{nk}.
\end{equation}
The second step utilizes Eq.(\ref{conjugate}). This quantity is interpreted as a weight associated with the potential measurement configuration $|a_j\rangle \rightarrow |s_m\rangle \rightarrow |s_n\rangle \rightarrow |a_k\rangle$. Suppose we do not perform actual measurement and inference information is not available, the probability of finding $S$ in a future measurement outcome can be calculated by summing $W_{jmnk}^{ASSA}$ from all applicable alternatives of measurement configurations. 

With this framework, the remaining task to calculate the probability is to correctly count the \textit{applicable} alternatives of measurement configuration. This task depends on the expected measurement outcome. For instance, suppose the expected outcome of an ideal measurement is event $|s_i\rangle$, i.e., measuring variable $q$ gives eigenvalue $q_i$. The probability of event $|s_i\rangle$ occurs, $p_i$, is proportional to the summation of $W_{jmnk}^{ASSA}$ from all the possible configurations related to $|s_i\rangle$. Mathematically, we select all $W_{jmnk}^{ASSA}$ with $m=n=i$, sum over index $j$ and $k$, and obtain the probability $p_i$.
\begin{equation}
\label{probability}
p_i \propto \sum_{j,k=0}^M (R^{SA}_{ij})^*R^{SA}_{ik}= |\sum_{j} R^{SA}_{ij}|^2.
\end{equation}
This leads to the definition of wave function $\varphi_i = \sum_{j} R_{ij}$, so that $p_i=|\varphi_i|^2$. The quantum state can be described either by the relational matrix $R$, or by a set of variables $\{\varphi_i\}$. The vector state of $S$ relative to $A$, is $|\psi\rangle_S^A = (\varphi_0, \varphi_1,\ldots, \varphi_N)^T$ where superscript $T$ is the transposition symbol. More specifically,
\begin{equation}
\label{WF}
|\psi\rangle_S^A = \sum_i\varphi_i |s_i\rangle  \quad \textrm{where } \varphi_i = \sum_{j} R_{ij}.
\end{equation}
The justification for the above definition is that the probability of finding $S$ in eigenvector $|s_i\rangle$ in future measurement can be calculated from it by defining a projection operator $\hat{P}_i=|s_i\rangle\langle s_i|$. Noted that $\{|s_i\rangle\}$ are orthogonal eigenbasis, the probability is rewritten as:
\begin{equation}
\label{probability3}
p_i=\langle\psi|\hat{P}_i|\psi\rangle = |\varphi_i|^2
\end{equation}

Eqs.(\ref{probability}) and (\ref{WF}) are introduced on the condition that there is no entanglement\footnote{See Section \ref{subsec:entanglement} for the definition of entanglement.} between quantum system $S$ and $A$. If there is entanglement between them, the summation in Eq.(\ref{probability}) over-counts the applicable alternatives of measurement configurations and should be modified accordingly. A more generic approach to describe the quantum state of $S$ is the reduced density matrix formulation, which is defined as
\begin{equation}
\label{reducedRho}
\rho_S=RR^\dag
\end{equation}
The probability $p_i$ is calculated using the projection operator $\hat{P}_i=|s_i\rangle\langle s_i|$
\begin{equation}
\label{indirectProb}
p_i = Tr_S(\hat{P}_i\hat{\rho}_S) = \sum_j|R_{ij}|^2.
\end{equation}

The effect of a quantum operation on the relational probability amplitude matrix can be expressed through an operator. Defined an operator $\hat{M}$ in Hilbert space ${\cal H}_S$ as $M_{ij} = \langle s_i|\Hat{M}|s_k\rangle$, The new relational probability amplitude matrix is obtained by
\begin{equation}
    \label{operator1}
    \begin{split}
    (R^{SA}_{new})_{ij} &= \sum_{k}M_{ik}(R^{SA}_{init})_{kj}, \quad \text{or} \\ R_{new} &= MR_{init}.
    \end{split}
\end{equation}
Consequently, the reduced density becomes,
\begin{equation}
    \label{operator3}
    \rho_{new} = R_{new}(R_{new})^\dag = M\rho_{init}M^\dag.
\end{equation}

\subsection{Entanglement Measure}
\label{subsec:entanglement}
The description of $S$ using the reduced density matrix $\rho_S$ is valid regardless there is entanglement between $S$ and $A$. To determine whether there is entanglement between $S$ and $A$, a parameter to characterize the entanglement measure should be introduced. There are many forms of entanglement measure~\cite{Nelson00, Horodecki}, the simplest one is the von Neumann entropy. Denote the eigenvalues of the reduced density matrix $\rho_S$ as $\{\lambda_i\}, i=0,\ldots, N$, the von Neumann entropy is defined as
\begin{equation}
    \label{vonNeumann}
    H(\rho_S)  = -\sum_i\lambda_i ln\lambda_i.
\end{equation}
A change in $H(\rho_S)$ implies there is change of entanglement between $S$ and $A$. Unless explicitly pointed out, we only consider the situation that $S$ is described by a single relational matrix $R$. In this case, the entanglement measure $E=H(\rho_S)$. Since $\rho_S=RR^\dag$, the entanglement measure is sometimes expressed as $H(R)$. Theorem 1 in Appendix \ref{Theorem1} provides a simple criteria to determine whether $H(R)=0$ based on the decomposition of $R_{ij}$.

$H(\rho_S)$ enables us to distinguish different quantum dynamics. Given a quantum system $S$ and its referencing apparatus $A$, there are two types of the dynamics between them. In the first type of dynamics, there is no physical interaction and no change in the entanglement measure between $S$ and $A$. $S$ is not necessarily isolated in the sense that it can still be entangled with $A$, but the entanglement measure remains unchanged. This type of dynamics is defined as \textit{time evolution}. In the second type of dynamics, there is a physical interaction and correlation information exchange between $S$ and $A$, i.e., the von Neumann entropy $H(\rho_S)$ changes. This type of dynamics is defined as \textit{quantum operation}. \textit{Quantum measurement} is a special type of quantum operation with a particular outcome. Whether the entanglement measure changes distinguishes a dynamic as either a time evolution or a quantum operation. 

Particularly, when $S$ is in an isolated state, its dynamics is governed by the Schr\"{o}dinger Equation~\cite{Yang2017}. One of the purposes of this paper is to provide the formulation of quantum operations when the entanglement measure between $S$ and $A$ changes.

\section{Quantum Measurement}
\label{sec:measurement}

\subsection{Expectation of A Measurement Theory}
The entanglement measure defined in Section \ref{subsec:entanglement} characterizes the quantum correlation between the measured system $S$ and the apparatus system $A$. The correlation enables the inference of measurement outcome. A change in entanglement measure implies change in the quantum correlation, consequently, change in the capability of inference. The capability of inference can be described by the mutual information variable, which is defined as~\cite{Hayashi15}
\begin{equation}
\label{MutualInfo}
    I(S, A) = H(\rho_S)+H(\rho_A)-H(\rho_{SA}).
\end{equation}
where $H(\rho)$ is the von Neumann entropy of a density matrix\footnote{However, there is speculation that quantum mutual information should be defined as $I(S, A) = H(\rho_S)-H(\rho_{SA})$, see remark in Ref.~\cite{Nielsen001}}. Mutual information is a quantity that measures the amount of information about $S$ through knowing information about $A$. For a pure bipartite state, $H(\rho_{SA})=0$ and $H(\rho_S)=H(\rho_A)$, thus $I(S,A)=2H(\rho_S)$, only differs from the Von Neumann entropy of the reduced density matrix of $S$ by a factor of 2. Thus, in this condition, it is equivalent to state that quantum operation is a process that alters the mutual information between $S$ and $A$. The term \textit{information exchange} used in the following text strictly refers the changes of mutual information.

Although the cause of information exchange is the physical interaction, the measurement theory in this paper does not aim to explain the detailed physical process on how $A$ records a particular outcome. Instead, the measurement theory just describes how the mutual information is transferred from one system to another. In the context of RQM, the goal is to describe how the relational probability amplitude matrix $R$ is transformed during measurement, and how mutual information is exchanged in the process.

Suppose the measurement is performed using apparatus $A$ and the initial correlation matrix is $R_0$. Although measurement dynamics involves information exchange between $S$ and $A$, the composite system $S+A$ is isolated, and can be described as a unitary process. This is Process 2 in the von Neumann measurement theory. The result is that the relational matrix $R_0$ is mapped to $R'$, denoted as $R_0\rightarrow R'$. $R'$ is then used by the intrinsic observer $\cal{O}_I$ to calculate the probability of a particular measurement outcome. Here the intrinsic observer is the one who reads the pointer variables of apparatus $A$, while any other observer who does not access the apparatus is called external observer. As pointed out in Ref.~\cite{Rovelli96}, if an external observer $\cal{O}_E$ only knows there is a measurement process occurred, but does not know the measurement outcome, his description of the measurement process is limited to $R_0\rightarrow R'$. On the other hand, the intrinsic observer, $\cal{O}_I$ who reads the pointer variable of $A$, knows the measurement outcome after the measurement finishes. This additional information on the exact measurement outcome allows $\cal{O}_I$ to infer the final quantum state of $S$. It results in another map $R'\rightarrow R^{\prime\prime}$. This is Process 1 in the von Neumann measurement theory. 

In summary, a measurement theory should describe how the relational matrix $R$ and the mutual information are changed during the measurement process. In this paper we also assume the description is with respect to a fixed QRF $F$. How the measurement formulation is transformed when switching QRFs is not in the scope of this paper but studies elsewhere~\cite{Yang2020}. We start the formulation with a simpler case that the $S+A$ composite system is initially in a product state.    

\subsection{Product Initial State}
\label{subsec:measProductState}
In Ref.~\cite{Yang2017}, it is shown that when the composite system $S+A$ is described by a relational probability amplitude matrix $R$ and assuming $S+A$ is in an isolated environment, it is mathematically equivalent to describe the composite system with a wave function,
\begin{equation}
\label{WFSA}
|\Psi\rangle = \sum_{ij}R_{ij}|s_i\rangle|a_j\rangle.
\end{equation}
Suppose $S$ and $A$ initially are unentangled, they can be described as a product state, $|\Psi_0\rangle_{SA}= \sum_{ij}R_{ij}|s_i\rangle|a_j\rangle$ where $R_{ij}=c_id_j$. This implies that $|\Psi_0\rangle_{SA}$ can be written as $|\psi_0\rangle_{S}|\phi_0\rangle_{A}$,  where $|\psi_0\rangle_{S}=\sum_ic_i|s_i\rangle$ and $|\phi_0\rangle_{A}=\sum_jd_j|a_j\rangle$. $S+A$ as a whole follows the Schr\"{o}dinger Equation. Since there is interaction between $S$ and $A$, the overall unitary operator cannot be decomposed to $\hat{U}_{SA} \ne \hat{U}_S\otimes \hat{U}_A$. Instead, $\hat{U}_{SA}$ can be decomposed such that it gives the following map
\begin{equation}
\label{map1}
\begin{split}
    |\Psi_1\rangle_{SA} & =\hat{U}_{SA}|\Psi_0\rangle_{SA} \\
    & = \hat{U}_{SA}|\psi_0\rangle_{S}|\phi_0\rangle_{A} \\
    & = \sum_{m}\hat{M}_m|\psi_0\rangle_S|a_m\rangle
\end{split}
\end{equation}
where the set of operators $\hat{M}_m$ satisfies the completeness condition $\sum_m\hat{M}_m\hat{M}_m^\dag = I$. Appendix A shows that such a decomposition always exists as long as the initial state is a product state. Substitute $|\psi_0\rangle=\sum_ic_i|s_i\rangle$ to Eq.(\ref{map1})
\begin{equation}
\label{map2}
|\Psi_1\rangle_{SA} = \sum_{im}(\sum_k(\hat{M}_m)_{ik}c_k)|s_i\rangle|a_m\rangle
\end{equation}
This gives the new relational matrix R' with element
\begin{equation}
\label{process2}
\begin{split}
    R^{\prime}_{im}& =\sum_k(\hat{M}_m)_{ik}c_k \\
    &=\sum_{jk}(\hat{M}_m)_{ik}(R_0)_{kj} = \sum_j(M_mR_0)_{ij}.
\end{split}
\end{equation}
$R^{\prime}_{im}$ cannot be decomposed into a format of $c_id_j$, which implies $H(R') \ne 0$ according to Theorem 1, and we cannot define a wave function for $S$. $R^{\prime}$ has now encoded the correlation between $S$ and $A$ and can be used to predict the probability of a possible measurement outcome.  At the end of the measurement, $\cal{O}_I$ who operates and reads the outcome of his apparatus $A$ knows the measurement outcome as $A$ ends up in a distinguishable pointer state $|a_m\rangle$. This allows $\cal{O}_I$ to infer exactly the resulting state of $S$. Since there is no additional interaction between $S$ and $A$, the process can be modeled as a local operator $I^S\otimes P_m^A$ where $P_m=|a_m\rangle\langle a_m|$. According to Theorem 2 in Appendix \ref{Theorem2}, the relational matrix is updated to $R^{\prime\prime}_m = I^SR^{\prime}(P_m^A)^T = R^{\prime}(P_m^A)^T$. Substituting $R^{\prime}$ obtained earlier, one has
\begin{equation}
\label{map3}
\begin{split}
(R^{\prime\prime}_m)_{ij} & = \sum_nR^{\prime}_{in}(P_m^T)_{nj} \\
& =\sum_n\sum_k(\hat{M}_n)_{ik}c_k\delta_{nm}\delta_{jm} \\
& =(\sum_k(\hat{M}_m)_{ik}c_k)(\delta_{jm})
\end{split}
\end{equation}
The last step shows that $(R^{\prime\prime}_m)_{ij}$ can be written as $c'_id'_j$ with $c'_i=\sum_k(\hat{M}_m)_{ik}c_k$ and $d'_j=\delta_{jm}$. From Theorem 1, $H(R^{\prime\prime}_m)=0$. Therefore, we can use Eq.(\ref{WF}) to calculate the wave function of $S$ corresponding to $|a_m\rangle$,
\begin{equation}
\label{measureWF}
\begin{split}
\varphi_i^m & = \sum_j(R^{\prime\prime}_m)_{ij}\\
& =\sum_k(\hat{M}_m)_{ik}c_k\sum_j\delta_{jm}\\
& =\sum_k(\hat{M}_m)_{ik}c_k
\end{split}
\end{equation}
Recall the initial state of $S$ is $|\psi_0\rangle_S =\sum_i c_i|s_i\rangle$, the resulting state vector for $S$, $|\psi_m\rangle = \sum_i\varphi_i^m |s_i\rangle$, can be written as $|\psi_m\rangle = \hat{M}_m|\psi_0\rangle_S$ without normalization. Applying the normalization factor, and omitting the subscript referring to $S$, one finally gets
\begin{equation}
\label{measureWF1}
|\psi_m\rangle = \frac{\hat{M}_m|\psi_0\rangle}{\sqrt{\langle\psi_0|\hat{M}_m^\dag \hat{M}_m|\psi_0\rangle}}
\end{equation}
The normalization factor is the probability of finding $A$ in the pointer state $|a_m\rangle$ after Process 2, i.e., the probability of measurement with outcome $m$. This can be verified by combining Eq.(\ref{partialProb}) in Appendix C and Eq.(\ref{map1}),
\begin{equation}
\label{measureProb}
\begin{split}
p_m & = \langle\Psi_1|I^S\otimes P_m^A|\Psi_1\rangle \\
& = \langle\psi_0|\hat{M}_m^\dag \hat{M}_m|\psi_0\rangle
\end{split}
\end{equation}
Due to the correlation in Eq.(\ref{map1}), the probability of finding $A$ in $|a_m\rangle$ is exactly the probability of inferring $S$ in the resulting state $|\psi_m\rangle$. If $\cal{O}_I$ repeats the same experiment many times, he shall find that the outcome $m$ occurs with a frequency of $p_m$, even though the outcome of a particular measurement is random.

Eqs.(\ref{measureWF1}) and (\ref{measureProb}) typically appear in textbooks as a postulate for quantum measurement~\cite{Nelson00}~\cite{Hayashi15}. In deriving these results, a mysterious ancillary system is introduced. The property of the ancillary system is traced out at the end to obtain Eq.(\ref{measureWF1}) and (\ref{measureProb}). As shown in this section, the ancillary system is nothing but the apparatus $A$. Its property can be traced out because the initial state is a product state, and at the end of the measurement, $S$ and $A$ are still in a product state. 

The last statement of the above paragraph needs more qualification. At the beginning of a measurement $H(R)=0$. At the end of the measurement $H(R^{\prime\prime}_m)=0$ as well. The entanglement measure appears to be the same at the beginning and at the end of the measurement. However, during the measurement process, $H(R)$ does not stay as a constant. This can be seen from Eq.(\ref{map1}). The correlation matrix $R^{\prime}_{im}=\sum_k(\hat{M}_m)_{ik}c_k$. It is not difficult to calculate\footnote{From Eq.(\ref{map1}), the reduced density operator of $S$ is $\hat{\rho}=Tr_A(|\Psi_1\rangle\langle\Psi_1|)=\sum_m\hat{M}_m|\psi_0\rangle\langle\psi_0|\hat{M}_m^\dag=\sum_m p_m|\psi_m\rangle\langle\psi_m|$. Thus $H(R^{\prime})=-Tr(\hat{\rho}ln(\hat{\rho}))=-\sum_mp_mln(p_m)$.} $H(R^{\prime})=-\sum_mp_mln(p_m)$. From Eq.(\ref{MutualInfo}) we can analyze the change of mutual information between $S$ and $A$. Initially $S$ and $A$ share no mutual information. In the initial phase of measurement, $S$ and $A$ interact and become entangled. Information from $S$ is encoded in $A$. The mutual information increases to $-2\sum_mp_mln(p_m)$. This allows an observer to infer probability of measurement outcome through $A$, but without knowing the exact measurement outcome. At the later phase of an ideal projection measurement, $A$ becomes disentangled with $S$ again and converges into a particular pointer state $|a_m\rangle$ with a probability $p_m$, this allows $\cal{O}_I$ to infer exactly which state $S$ is in. When the measurement ends, $S$ and $A$ share no mutual information again. During the measurement, the mutual information is changed as $0\rightarrow -2\sum_mp_mln(p_m)\rightarrow 0$. 

The increase of mutual information in the first arrow is described as a unitary process of the composite system, and the decrease of mutual information in the second arrow is described by a projection operator. The update of the relational matrix from $R^{\prime}$ to $R^{\prime\prime}$ was perceived as ``wave function collapse" in the Copenhagen Interpretation. However, in this paper this update is not associated with a physical reality change. Instead, it is interpreted as change of description of the relational matrix due to the fact that $\cal{O}_I$ knows the exact measurement outcome. An external observer $\cal{O}_E$ does not know the measurement outcome and therefore still describes $S$ with $R^{\prime}$. We see that even though both $\cal{O}_I$ and $\cal{O}_I$ describe $S$ through a relational matrix, the relational matrix itself is relative, as pointed out in the introduction section. $\cal{O}_E$ can obtain the measurement outcome through communication with $\cal{O}_I$. But this means there is a physical interaction between $A'$ and $A$. An interaction between $A'$ and $A$ disqualifies $\cal{O}_E$ to describe the composite system $S+A$ as a unitary time evolution. Thus Process 1 cannot be described as a unitary process by either $\cal{O}_I$ or $\cal{O}_E$. In other words, Process 1 cannot be described by the Schr\"{o}dinger Equation. One of the preconditions for applying Schr\"{o}dinger Equation is that there should have no information exchange between the observed system and the reference apparatus. But for an observer to know the exact measurement outcome, such information exchange is unavoidable. 

As mentioned earlier, measurement theory is to develop a physical model that describes how mutual information is exchanged during measurement. The detailed physical process of interaction is not explained here. For example, after the $S$ and $A$ become entangled, one must assume there is no further interaction between $S$ and $A$ in order to model the process with the local projection operator $I^S\otimes P_m^A$. It may be just an approximation. Ref.~\cite{Allah13} provides tremendous amount of physical details to describe this process. The measurement process goes through several sub-processes such as registration, truncation, decoherence, and the emergence of a unique outcome that is interpreted using quantum statistics mechanics~\cite{Allah13}. It is of great interest to find out the physical details on the measurement process, but the primary interest of the measurement theory developed here is how the relational matrix $R$ and the mutual information are changed during the measurement process.

\subsection{Entangled Initial State}
When $S$ and $A$ are initially entangled, $A$ already has some level of correlation with $S$. In a sense that $A$ has already measured $S$ since the information of $A$ can be used to infer information of $S$. One may ask what the goal of subsequent measurement is in this case. In the situation that $S$ and $A$ are initially in product state, an operation involving interaction between $S$ and $A$ increases the mutual information, thus allowing $A$ to infer information of $S$. Similarly, in the case when $S$ and $A$ are initially entangled, the goal of the measurement can be further increasing the mutual information. After more mutual information is encoded in $A$, a subsequent projection operation can be applied so that $A$ evolves to a unique distinguishable pointer state. Since the mutual information is defined as $I(S, A) = H(\rho_S)+H(\rho_A)-H(\rho_{SA})$, the maximum mutual information for a pure bipartite state is $I_{max}=2lnN$ where $N$ is the rank of matrix $R$. We can define the amount of unmeasured mutual information as
\begin{equation}
\label{MI}
I_u(S, A) = 2lnN - I(S, A).
\end{equation}
Thus, the goal of measurement is to minimize $I_u(S, A)$. 

Alternatively, the goal of measurement can be set as to alter the probability distribution $\{p_m\}$ such that the probability to find $S$ in a particular state is adjusted as desired, or such that the expectation value of an observable of $S$ matches a desired value. We will discuss both cases in this section.

Denote the initial entangled state for $S+A$ as $|\Psi_0\rangle_{SA} = \sum_{ij}R_{ij}|s_i\rangle|a_j\rangle$. The interaction between $S$ and $A$ is still described as a unitary operation over the whole $S+A$ composite system, $|\Psi_1\rangle_{SA}=\hat{U}_{SA}|\Psi_0\rangle_{SA}$. The relational matrix $R'$ is 
\begin{equation}
    \label{R1}
    \begin{split}
    R^\prime_{ij}& = \langle s_i|\langle a_j|\hat{U}_{SA}|\Psi_0\rangle \\
    & = \sum_{kl}R_{kl}\langle s_i|\langle a_j|\hat{U}_{SA}|s_k\rangle|a_l\rangle.
    \end{split}
\end{equation}
Then $A$ is projected to a particular state $|a_m\rangle$. Similar to the approach in deriving Eq.(\ref{map3}), the relational matrix is updated to $R^{\prime\prime}_m = I^SR^{\prime}(P_m^A)^T$ where $P_m^A = |\phi_m\rangle\langle\phi_m|$, we get
\begin{equation}
\label{map4}
\begin{split}
(R^{\prime\prime}_m)_{ij} & = \sum_nR^{\prime}_{in}(P_m^T)_{nj} \\
& =\sum_{n}\langle s_i|\langle a_n|\hat{U}_{SA}|\Psi_0\rangle\langle a_j|\phi_m\rangle\langle\phi_m|a_n\rangle \\
&= \sum_n\langle\phi_m|a_n\rangle\langle a_n|\langle s_i|\hat{U}_{SA}|\Psi_0\rangle\langle a_j|\phi_m\rangle \\
& = \langle s_i|\langle\phi_m|\hat{U}_{SA}|\Psi_0\rangle\langle a_j|\phi_m\rangle
\end{split}
\end{equation}
where the property $\sum_n |a_n\rangle\langle a_n|=I$ is applied in the third step. Since $(R^{\prime\prime}_m)_{ij}$ can be written as product of two terms with index $i$ and $j$ separated, according to Theorem 1, $H(R)=0$. We can use Eq.(\ref{WF}) to calculate the wave function of $S$ associated with outcome $|a_m\rangle$
\begin{equation}
\label{measureWF2}
\begin{split}
\varphi_i^m & = \sum_j(R^{\prime\prime}_m)_{ij} \\
& = \langle s_i|\langle\phi_m|\hat{U}_{SA}|\Psi_0\rangle \sum_j \langle a_j|\phi_m\rangle \\
& = d_m \langle s_i|\langle \phi_m|\hat{U}_{SA}|\Psi_0\rangle
\end{split}
\end{equation}
where $d_m=\sum_j \langle a_j|\phi_m\rangle$ is a normalization constant.  The probability of finding measurement outcome associated with $|a_m\rangle$ is given by Eq.(\ref{partialProb}) in Appendix C,
\begin{equation}
\label{measureProb2}
\begin{split}
p'_m &= \langle\Psi_1|I^S\otimes P_m^A|\Psi_1\rangle \\
& = \sum_i|\langle s_i|\langle\phi_m|\hat{U}|\Psi_0\rangle|^2 \\
& = \langle\Psi_0|\hat{U}^\dag|\phi_m\rangle\langle\phi_m|\hat{U}|\Psi_0\rangle.
\end{split}
\end{equation}
The resulting state vector of $S$ before normalization is
\begin{equation}
\label{measureWF3}
\begin{split}
|\psi_m\rangle & =\sum_i\varphi_i^m|s_i\rangle\\
& = d_m\sum_i |s_i\rangle\langle s_i|\langle \phi_m|\hat{U}_{SA}|\Psi_0\rangle \\
& = d_m \langle \phi_m|\hat{U}_{SA}|\Psi_0\rangle
\end{split}
\end{equation}
Normalization requires that $d_m=1/\sqrt{p'_m}$. In order to simplify Eq.(\ref{measureWF3}), we rewrite the initial entangled bipartite state using the Schmidt decomposition $|\Psi_0\rangle = (U_S\otimes V_A)\sum_i\lambda_i|\tilde{s}_i\rangle|\tilde{a}_i\rangle$ where $U_S\otimes V_A$ is a local unitary transformation. $\lambda_i$ is the Schmidt coefficient, which essentially is the eigenvalue of the relational matrix $R$. This gives
\begin{equation}
\label{measureWF4}
\begin{split}
|\psi_m\rangle & =\frac{1}{\sqrt{p'_m}} \sum_i\langle \phi_m|\hat{U}_{SA}(U_S\otimes V_A)|\tilde{s}_i\rangle|\tilde{a}_i\rangle \\
& = \frac{1}{\sqrt{p'_m}} \sum_i \hat{M}_{mi}|\tilde{s}_i\rangle
\end{split}
\end{equation}
where $\hat{M}_{mi}=\lambda_i\langle\phi_m|\hat{U}_{SA}(U_S\otimes V_A)|\tilde{a}_i\rangle$. Note that $\hat{M}_{mi}$ depends on the initial state itself, Eq.(\ref{measureWF4}) is not a simple form. If $|\Psi_0\rangle$ is a product state, $\lambda_0=1$ and $\lambda_i=0$ for $i>0$, Eq.(\ref{measureWF4}) is reduced to Eq.(\ref{measureWF1}). Given that $\sum_m\hat{P}_m = I$, it is easy to verify the completeness property of $\hat{M}_{mi}$
\begin{equation}
\label{MeasOperators}
\begin{split}
&\sum_m \hat{M}_{mi}^\dag\hat{M}_{mj} = \delta_{ij}\lambda_i^2I_S \\ &\sum_m\sum_{ij}\hat{M}_{mi}^\dag\hat{M}_{mj} = I_S.
\end{split}
\end{equation}
Since $\langle\psi_m|\psi_m\rangle=1$, from Eq.(\ref{measureWF4}) one gets $p'_m =\sum_{ij}\langle \tilde{s}_i|\hat{M}_{mi}^\dag\hat{M}_{mj}|\tilde{s}_j\rangle$. It follows from Eq.(\ref{MeasOperators}) that $\sum_m p'_m= \sum_i\lambda_i^2=1$. From the expression for $p'_m$, the mutual information after the unitary operation can be calculated as $I'(S,A)=-2\sum_mp'_mln(p'_m)$. On the other hand, the initial mutual information $I(S,A)=-2\sum_i|\lambda_i|^2ln(|\lambda_i|^2)$. If the goal of measurement is to increase the mutual information, one wishes to find a unitary operator $\hat{U}$ such that $I'(S,A)> I(S,A)$, that is,
\begin{equation}
\label{mutualInfoIneq}
\sum_mp'_mln(p'_m) < \sum_i|\lambda_i|^2ln(|\lambda_i|^2).
\end{equation}

On the other hand, if the goal of measurement is not necessarily to increase the mutual information, but to increase the probability that $S$ is in a state inferred by $A$ being in the pointer state $|a_m\rangle$. The initial probability before measurement operation is $p_m=\sum_i|R_{im}|^2$. After measurement operation, we want $p'_m > p_m$. This means the goal of measurement is to find a unitary operator $\hat{U}_{SA}$ such that 
\begin{equation}
\label{mutualInfoIneq2}
\sum_{ij}\langle \tilde{s}_i|\hat{M}_{mi}^\dag\hat{M}_{mj}|\tilde{s}_j\rangle > \sum_i|R_{im}|^2.
\end{equation}
Neither Eq.(\ref{mutualInfoIneq}) nor Eq.(\ref{mutualInfoIneq2}) is simple to solve. It is not clear that for a given initial correlation matrix $R$, a unitary operator $\hat{U}_{SA}$ that satisfies either Eq.(\ref{mutualInfoIneq}) or Eq.(\ref{mutualInfoIneq2}) always exists. This is an open topic for future research.

The quantum measurement theory developed in the RQM context is equivalent to the Open Quantum System theory, if we replace the environment system in OQS with the reference apparatus system in this formulation. The details of the equivalency is left in Appendix E.

\section{General Quantum Operation}
\label{sec:generalOp}
In Section \ref{sec:measurement} we only consider the selective measurement. At the end of the selective measurement operations, the apparatus $A$ is in a definite state, and $S$ and $A$ are in a product composite state. There is other type of quantum operation where at the end of the operation, $S$ and $A$ are in an entangled state and there are still multiple possible outcomes. This is the non-selective measurement~\cite{Breuer07}. For instance, the composite system $S+A$ can go through the interaction characterized by an operator $\Lambda_{SA} \ne \hat{U}_S\otimes \hat{U}_A$ and there is no further projection operation. $S$ and $A$ are entangled at the end of the operation.

A more general global linear map on a bipartite system can be decomposed to $\Lambda_{SA}=\sum_k\alpha_k\hat{B}_k\otimes \hat{C}_k$ where $\hat{B}_k$ is local operator to $S$ and $\hat{C}_k$ is local operator to $A$~\cite{Nelson00}. $\Lambda_{SA}$ is a general operation in the sense that $\hat{B}_k$ or $\hat{C}_k$ are not necessarily project operators, and the resulting $S+A$ can be in a product state or an entangled state. It is convenient to re-express the relational matrix by introduce a linear operator $\hat{R}=\sum_{ij}R_{ij}|s_i\rangle\langle a_j|$. According to Theorem 2, the operation of $\Lambda_{SA}$ on $S+A$ transforms the initial relational operator $\hat{R}_0$ to
\begin{equation}
\label{entangledR3}
    \hat{R} = \Lambda_{SA}(\hat{R}_0) = \sum_k\alpha_k\hat{B}_k\hat{R}_0\hat{C}_k^T
\end{equation}
The reduced density operator for $S$ after the general quantum operation is
\begin{equation}
\label{entMeasR4}
\hat{\rho}_S = \hat{R}\hat{R}^\dag  = \sum_{kl}\alpha_k\alpha_l^* \hat{B}_k\hat{R}_0(\hat{C}_l^\dag\hat{C}_k)^T\hat{R}_0\hat{B}_l^\dag
\end{equation}
Suppose the initial composite state of $S+A$ is $|\Psi_0\rangle =\sum_i\lambda_i|\tilde{s}_i\rangle|\tilde{a}_i\rangle$, the relational operator $\hat{R}_0$ can be expressed as $\hat{R}_0=\sum_i\lambda_i|\tilde{s}_i\rangle\langle\tilde{a}_i|$. Substitute this into Eq.(\ref{entMeasR4}),
\begin{equation}
\label{entMeasR5}
\begin{split}
\hat{\rho}_S & = \hat{R}\hat{R}^\dag \\
& = \sum_{ijkl}\lambda_i\lambda_j\alpha_k\alpha_l^* \hat{B}_k|\tilde{s}_i\rangle\langle\tilde{a}_i|(\hat{C}_l^\dag\hat{C}_k)^T|\tilde{a}_j\rangle\langle\tilde{s}_j|\hat{B}_l^\dag \\
&=\sum_{ijkl}(\lambda_i\lambda_j\alpha_k\alpha_l^*\langle\tilde{a}_j|\hat{C}_l^\dag\hat{C}_k|\tilde{a}_i\rangle)\hat{B}_k|\tilde{s}_i\rangle\langle\tilde{s}_j|\hat{B}_l^\dag.
\end{split}
\end{equation}
If the local operator $\hat{C}$ is a unit operator, $\Lambda_{SA}=\sum_k\alpha_k\hat{B}_k\otimes I_A = \Lambda_S \otimes I_A$. This means $\Lambda_{SA}$ only operates on $S$ and has no impact on $A$. Eq.(\ref{entMeasR5}) becomes
\begin{equation}
\label{entMeasR6}
    \hat{\rho}_S = \Lambda_S\hat{\rho}_0 \Lambda_S^\dag
\end{equation}
where $\hat{\rho}_0=\sum_i\lambda_i^2|\tilde{s}_i\rangle\langle\tilde{s}_i|$ is the initial density operator for $S$. However, if $\Lambda_{SA} \ne \Lambda_S \otimes I_A$, Eq.(\ref{entMeasR6}) does not hold in general.

Eq.(\ref{entMeasR5}) can be derived through the partial trace approach as well. The initial density operator of the composite system is $\rho_{SA}=|\Psi_0\rangle\langle \Psi_0|=\sum_{ij}\lambda_i\lambda_j|\tilde{s}_i\rangle|\tilde{a}_i\rangle\langle\tilde{s}_j|\langle\tilde{a}_j|$. After applying the general operation $\Lambda_{SA}$, the reduced density operator of $S$ is
\begin{equation}
\label{entMeasR7}
\begin{split}
\hat{\rho}_S & = Tr_A(\Lambda_{SA}\rho_{SA}\Lambda_{SA}^\dag) \\
& = Tr_A(\sum_{ijkl}\lambda_i\lambda_j\alpha_k\alpha_l^* \hat{B}_k|\tilde{s}_i\rangle\langle\tilde{s}_j|\hat{B}_l^\dag \otimes \hat{C}_k|\tilde{a}_i\rangle\langle\tilde{a}_j|\hat{C}_l^\dag) \\
&= \sum_{ijkl}\lambda_i\lambda_j\alpha_k\alpha_l^* \hat{B}_k|\tilde{s}_i\rangle\langle\tilde{s}_j|\hat{B}_l^\dag \{Tr_A(\hat{C}_k|\tilde{a}_i\rangle\langle\tilde{a}_j|\hat{C}_l^\dag)\} \\
&= \sum_{ijkl}\lambda_i\lambda_j\alpha_k\alpha_l^*(\langle\tilde{a}_j|\hat{C}_l^\dag\hat{C}_k|\tilde{a}_i\rangle) \hat{B}_k|\tilde{s}_i\rangle\langle\tilde{s}_j|\hat{B}_l^\dag
\end{split}
\end{equation}
which is the same as Eq.(\ref{entMeasR5}).

An application of Eq.(\ref{entMeasR5}) is briefly described as following. In Section \ref{sec:measurement}, we have been assuming the measuring apparatus is $A$. The initial interaction between $S$ and $A$ is described as a unitary operation on $S+A$. This is equivalent to the case that the general map $\Lambda_{SA}$ is a unitary operator and can be decomposed according to (\ref{map1}). However, the measurement of $S$ can be performed using another apparatus $A'$. In this case, $A'$ shall interact with either $S$ or the $S+A$ composite system. The general map $\Lambda_{SA}$ is not a unitary operator anymore. Instead, it can be considered as a quantum operation decomposed from a unitary operator for the $S+A+A'$ composite system. If $A'$ interacts with both $S$ and $A$, the resulting reduced density operator of $S$ is given by Eq.(\ref{entMeasR5}). If the apparatus $A'$ only interacts with $S$ and has no impact on $A$, $\hat{C}$ is a unit operator, and the resulting reduced density operator of $S$ is given by Eq.(\ref{entMeasR6}).

Eq.(\ref{entangledR3}) is the most general form of equation describing different types of dynamics between $S$ and $A$, depending on how the map $\Lambda_{SA}$ is decomposed. If $\Lambda_{SA} = \hat{U}_S\otimes \hat{U}_A$, it results in the Schr\"{o}dinger Equation~\cite{Yang2017}. If $\Lambda_{SA}$ is a unitary operator but decomposed according to Eq.(\ref{map1}), it describes process 2 of the von-Neumann measurement process. If $\Lambda_{SA}=I^S\otimes P_m^A$ where $P_m=|a_m\rangle\langle a_m|$, it describes the process 1 of the measurement process. Lastly, the most general decomposition of $\Lambda_{SA}$ gives Eq.(\ref{entangledR3}). One logical conclusion is that Schr\"{o}dinger Equation cannot describe all these quantum dynamics, particularly, cannot described the process 1 in the measurement process, as discussed in the previous section. 

\section{Explicit v.s. Implicit Relativity}
\label{criteria}
In the relational formulation of quantum mechanics, even though a quantum system $S$ should be described relative to a reference system $A$, there are mathematical tools that provide equivalent descriptions without explicitly calling out the reference system $A$. When $S$ and $A$ are unentangled, $S$ can be described by a wave function defined in Eq.(\ref{WF}). When $S$ and $A$ are entangled, $S$ is described by a reduced density matrix that traces out the information of $A$. On the other hand, the measurement theory developed here indicates the observer must be called out explicitly when describing a measurement process. We wish to obtain a cohesive conclusion on when the reference system must be explicitly called out.

If $S$ is unentangled with any other quantum system, it can be described with a wave function defined as Eq.(\ref{WF}). Supposed there are two different observers with their own pieces of apparatus $A$ and $A'$, and the relational matrices are $R$ and $R'$, respectively. The wave function for $S$ is $\varphi_i=\sum_jR_{ij}$ relative to $A$, and $\varphi'_i=\sum_jR'_{ij}$ relative to $A'$. However, since there is no entanglement between $S$ and $A$, or between $S$ and $A'$, according to Theorem 1, $R_{ij}$ can be decomposed as $R_{ij}=c_id_j$. Therefore, $\varphi_i=c_i\sum_jd_j=dc_i$, where $d=\sum_jd_j$ is just a constant. If $R'_{ij}$ is decomposed as $R_{ij}=c_id'_j$, $\varphi'_i=c_i\sum_jd'_j=d'c_i$. Thus, $\varphi_i$ and $\varphi'_i$ are different only by an unimportant constant. The description of $S$ relative to $A'$ is equivalent to the description relative to $A$. In this case, there is no negative consequence\footnote{Note that if $R_{ij}=c'_id'_j$, $\varphi'=d'c'_i \ne \varphi$. We do not prove $R_{ij}=c_id'_j$ here but only show it is a possible decomposition. The proof is not necessary because our goal is to show it is \textit{possible} to describe $S$ without calling out $A$.} to describe $S$ without calling out $A$. 

If $S$ is entangled with another system $A$ but the $S+A$ composite system is unentangled with any other system, based on the same reasoning in the previous paragraph, the $S+A$ composite system can be described without calling out the reference system. The state vector of the composite system is~\cite{Yang2017}
\begin{equation}
\label{WF3}
\begin{split}
|\Psi\rangle & = \sum_m \varphi_m|m\rangle = \sum_{ij} \varphi_{ij} |s_i a_j\rangle \\
& = \sum_{ij}R_{ij}|s_i\rangle|a_j\rangle.
\end{split}
\end{equation}
$S$ itself is described by the reduced density matrix, 
\begin{equation}
    \label{density}
    \begin{split}
    \hat{\rho}_S & =Tr_A|\Psi\rangle\langle\Psi| = \sum_{ii'}(\sum_kR_{ik}R^*_{i'k})|s_i\rangle\langle s_{i'}| \\
    & =\sum_{ii'}(RR^\dag)_{ii'}|s_i\rangle\langle s_{i'}|.
    \end{split}
\end{equation}
Given $|\Psi\rangle$ can be described without calling out the reference system and the reduced density matrix $\rho_S$ is derived from $|\Psi\rangle$, it is logical to deduce that $\rho_S$ can be described without calling out the reference system either.

Since time evolution is defined as a quantum process that there is no change of entanglement measure between $S$ and any other system, the argument presented above holds true for any given moment during time evolution. Supposed the time evolution Hamiltonian operator is known to any observer, there is no need to call out explicitly the reference system in the description. 

Quantum measurement process, on the other hand, is different. The measurement process comprises two sub-processes. In process 2, the composite system $S+A$ can be described as a unitary process. There is information exchange between $S$ and $A$, but $S+A$ as a whole does not exchange information with other system. The intrinsic observer, $\cal{O}_I$, who is associated with $A$, describes Process 2 according to Eq.(\ref{process2}); An external observer, $\cal{O}_E$, who is associated with $A'$, describes the same process according to Eq.(\ref{map1}). On the condition that both $\cal{O}_I$ and $\cal{O}_E$ have the same information of the total Hamiltonian operator, they can have the equivalent descriptions on $S$. As for Process 1, it is modeled as a projection operator and is not a unitary process. $\cal{O}_I$, who is associated with $A$ and reads the outcome from $A$, gains additional information compared with other observers $\cal{O}_E$, who do not know the exact the measurement outcome. This is not a unitary process\footnote{One may argue that if we include the observer herself into the composite system $S+A+O$, the entire composite system can be treated as an isolated system and described as going through a unitary process. However the inclusion of $O$ into the described system means there is yet another apparatus is involved that can measure the $S+A+O$ composite system, thus it means a change of observer by definition. Furthermore, such approach still cannot explain why a single outcome is selected at the end of measurement. In fact, the decoherence theory follows such reasoning by considering $O$ as environment of the apparatus. But the decoherence theory does not explain why at the end of a measurement a single outcome is singled out from all possible outcomes after decoherence takes place. The Quantum Bayesian Theory models Process 1 as a probability update after additional data is collected. Obviously, the Bayesian probability update is not a unitary process.} relative to either $\cal{O}_I$ or $\cal{O}_E$. We conclude that a measurement process, comprising both Processes 1 and 2, must be described by calling out the reference system explicitly. A quantum process that must be described by calling out the reference system is defined to be explicitly relative. On the other hand, a quantum process that can be described without calling out the reference system is implicitly relative. We can summary our conclusions with the following statement:
\begin{displayquote}
\textit{For a given system and the initial condition, the time evolution process is implicitly relative, while the measurement process is explicitly relative.}
\end{displayquote}
By definition, in a measurement process the entropy of the observed system $S$, $H(R)$, is changed. A change of entropy of $S$ implies there is information exchange between $S$ and another system. Thus, the above statement can be restated as
\begin{displayquote}
\textit{Given a system and the initial condition, if a quantum process induces information exchange from $S$, the process is explicitly relative.}
\end{displayquote}
The statement of ``given a system and the initial condition" is important. In the Process 2 of a quantum measurement process, $S$ interacts with $A$ during a measurement. There is entropy change for $S$, thus the process is explicitly relative, i.e., relative to $\cal{O}_I$. If we change the boundary of the system to $S+A$, we need another apparatus $A'$. This means the observer is changed to $\cal{O}_E$. $\cal{O}_E$ can derive equivalent description of $S$ compared to that relative to $\cal{O}_I$, as shown in Eqs.(\ref{map1}) and (\ref{process2}). This is on the condition that $\cal{O}_E$ knows the total Hamiltonian operator of $S+A$. However, $\cal{O}_E$ and $\cal{O}_I$ may not always share the same information, as seen in Process 1.

Suppose system $S$ comprises two subsystems $S1$ and $S2$, and the two subsystems are space-like separated. If $S1+S2$ do not interact with other system, the entropy of $S1+S2$ is unchanged. Even in the case $S1$ and $S2$ interacts with each other and the reduced entropy of $S1$ or $S2$ changes, $S1+S2$ as a whole undergoes time evolution and the process is implicitly relative. However, if either $S1$ or $S2$ interacts with another system outside the composite system, the process is explicitly relative. This is at the heart of the EPR paradox.

\section{EPR}
\label{subsec:EPR}
\subsection{Hidden Assumptions}
In traditional quantum mechanics, since the reference system is not typically called out in traditional quantum mechanics, there are assumptions on the reference apparatus that are not obvious. Two of such hidden assumptions are:
\begin{enumerate}
    \item An unentangled reference apparatus always exists regardless the composition of the observed system $S$. For instance, $S$ can be as large as the Universe.
    \item Suppose $S$ comprises multiple subsystems and these subsystems are space-like separated. When $A$ measures a subsystem of $S$, the observer knows the measurement result instantaneously regardless where the observer locates.  
\end{enumerate}
Let us call such an observer who knows the measurement result instantaneously as a Super Observer $\cal{O}_S$. Because $\cal{O}_S$ always exists and knows the changes of the relational matrix $R$ instantaneously, one can choose the apparatus associated with $\cal{O}_S$ as an absolute reference. A quantum state can then be described as an absolute state. The assumption that there exists a Super Observer enables the notion of absolute state for a quantum system~\footnote{The role of a privileged observer is also proposed in Ref~\cite{Rovelli96} in the effort to resolve the classical and quantum world separation issue in the Copenhagen Interpretation. This privileged system is similar to the super observer (or, super-apparatus) contains collection of ``all the macroscopic objects around us"~\cite{Rovelli96}.}. In most of physical processes where $S$ is an isolated system and the locations of its subsystems are sufficiently close, an observer-independent state will not lead to paradox. Mathematically it is more convenient and elegant to describe a quantum state as observer-independent. However, when a quantum system comprises two entangled subsystems and the two subsystems are remotely separated, the view of $\cal{O}_S$ can lead to the paradox described in the EPR paper~\cite{EPR}. Ref.~\cite{Rovelli07} had already provided a thorough analysis of the EPR paradox in the RQM context. This analysis presented in this paper is in general consistent with the argument in Ref.~\cite{Rovelli07}, however, we bring more insights on the role of a hidden Super Observer in the EPR argument, and explore the implication of resolution from information exchange perspective.

\subsection{EPR Argument}
The EPR argument is briefly reviewed as following. Assuming two systems $\alpha$ and $\beta$ are initially in the same physical location and interact for a period of time. They become entangled and then move away from each other with a space-like separation. We will adopt Bohm's version of the EPR argument by assuming $\alpha$ and $\beta$ are two spin half particles. The quantum state of the composite system can be decomposed based on the up and down eigenstates along the $z$ direction, or left and right eigenstates along the $x$ direction\footnote{Note that $|\alpha_l\rangle=\frac{1}{\sqrt{2}}(|\alpha_u\rangle+|\alpha_d\rangle)$ and $|\alpha_r\rangle=\frac{1}{\sqrt{2}}(|\alpha_u\rangle-|\alpha_d\rangle)$ }:
\begin{equation}
\label{EPR}
\begin{split}
|\Psi\rangle & = \frac{1}{\sqrt{2}}(|\alpha_u\rangle|\beta_u\rangle - |\alpha_d\rangle|\beta_d\rangle) \\
& = \frac{1}{\sqrt{2}}(|\alpha_l\rangle|\beta_l\rangle - |\alpha_r\rangle|\beta_r\rangle).
\end{split}
\end{equation}
In the context of RQM, Eq.(\ref{EPR}) assumes there is an observer, Alice, with apparatus $A$ that can measure $\alpha$ and $\beta$, and $A$ is unentangled with the composite system. The issue here is that after $\alpha$ and $\beta$ are remotely separated, there is no apparatus that can perform measurement on $\alpha$ and $\beta$ at the same time. Alice needs to perform measurement on $\alpha$ or $\beta$ once at a time. Alternatively, there can be two local observers, Alice with apparatus $A$ and located with particle $\alpha$, and Bob with apparatus $B$ and located with particle $\beta$, to perform the measurements at the same time.

The EPR paper then proposed a definition of realism as following,
\begin{displayquote}
\textit{If, without in any way disturbing a system, we can predict with certainty the value of a physical quantity, then there exists an element of physical reality corresponding to this physical quantity.}~\cite{EPR}
\end{displayquote}
Now Alice performs a measurement on $\alpha$ and supposed the outcome is spin up. Traditional quantum mechanics states that wave function vector $|\Psi\rangle$ collapses to $|\alpha_u\rangle|\beta_u\rangle$, thus $\beta$ is deterministically in the spin up state after Alice's measurement. If instead Alice performs a measurement along $x$ axis and finds $\alpha$ is in the spin left eigenstate, $\beta$ is deterministically in the left eigenstate after the measurement. A measurement in the location where $\alpha$ is does not cause a state change for $\beta$ that is space-like separated, otherwise it violates the principle of locality demanded by special relativity. Since one can predict the spin of $\beta$ in both $z$ and $x$ directions without disturbing it, by the above definition of realism, $\beta$ can simultaneously have elements of reality for the spin properties in both $z$ and $x$ directions. Denoting these properties as eigenvalues of operators $\sigma_z$ and $\sigma_x$. However, $\sigma_z$ and $\sigma_x$ are non-commutative. According to Heisenberg Uncertainty Principle, $\beta$ cannot simultaneously have definite eigenvalues for $\sigma_z$ and $\sigma_x$. Therefore, there are elements of physical reality of $\beta$ that the quantum mechanics cannot describe. This leads to the conclusion that quantum mechanics is an incomplete theory. 

The issue here is that the definition of realism assumes the element of physical reality is observer-independent. It assumes the measurement of Alice on $\alpha$ reveals a physical reality that is observer-independent, and Bob at a remote location knows the same physical reality instantaneously. But both Alice and Bob are local observers, such definition is not operational to them, unless faster that light interaction is permitted. If, however, there is another observer, Charles, who always know the state of $\alpha$ and $\beta$ at the same time, any measurement on either $\alpha$ or $\beta$ is known to Charles instantaneously. With the help of Charles, the definition of absolute physical reality is operational. However, Charles is a super observer according to our definition. Such an observer is imaginary, although we unintentionally assume he always exists, and we build physical concepts with such assumption. It is the assumption that there exists a Super Observer that allows the definition of absolute element of physical realism. According to the criteria presented in Section \ref{criteria}, for the given composite system $\alpha + \beta$, the process of Alice's measurement on $\alpha$ extracts information the composite system. Therefore it must be described by calling out the observer explicitly. 

\subsection{Resolution}
The original definition of an element of physical realism depends on a Super Observer and is not operational. Instead, the definition should be modified as following,
\begin{displayquote}
\textit{If, without in any way disturbing a system, a local observer can predict with certainty the value of a physical quantity, then there exists an element of physical reality corresponding to this physical quantity, relative this local observer.}
\end{displayquote}
With this modified definition, let's proceed the EPR reasoning to see whether it leads to the conclusion of incompleteness of quantum mechanics. If Alice performs a measurement on $\alpha$ along the $z$ direction and the outcome is spin up, the wave function after measurement is updated to $|\alpha_u\rangle|\beta_u\rangle$. This just means that $\beta$ is in the spin up state according to Alice. The element of physical reality is true only relative to Alice. From Bob's perspective, before he knows the Alice's measurement result, he still views the composite system in the original state, no quantum event happened yet. In other words, Bob still predicts that future measurement on $\beta$ will find it is in spin up state with fifty percentage of chance. At this point, both observers are out of synchronization on the relational information of the two particles, thus give different descriptions of particle $\beta$. Alice's description is ``Given the condition that $\alpha$ is in the spin up state, $\beta$ is in spin up state with unit probability", while Bob's description is still ``$\beta$ is in spin up state with fifty percent of chance". Note that Alice's description contains a new condition ``$\alpha$ is in the spin up state" while Bob's description doesn't contain such condition. Both descriptions are valid. To verify the physical description Alice obtained on particle $\beta$ after measuring particle $\alpha$, Alice can travel to Bob's location to perform a measurement, or can send the measurement result to Bob and ask Bob to perform a measurement. Suppose Alice sends the measurement outcome to Bob. Bob updates the wave function accordingly to $|\alpha_u\rangle|\beta_u\rangle$, same as the wave function relative to Alice. He now can confirm the physical reality that $\beta$ is in spin up state $|\beta_u\rangle$ with unit probability. However, in this state, he cannot predict deterministically that $\beta$ is in spin left or right, since $|\beta_u\rangle = \frac{1}{\sqrt{2}}(|\beta_l\rangle + |\beta_r\rangle)$. Similarly, if Alice performs a measurement on $\alpha$ along the $x$ direction and the outcome is spin left, $\beta$ is deterministically in the spin left state relative to Alice, but nothing happened from Bob's perspective. If Alice sends the measurement result to Bob, Bob updates the wave function accordingly to $|\alpha_l\rangle|\beta_l\rangle$. He now can confirm the physical reality that $\beta$ is spin left state $|\beta_l\rangle$. However, in this state, he cannot predict deterministically that $\beta$ is in spin up or down. Since Alice cannot perform measurement on $\alpha$ along $z$ and $x$ directions in the same time, Bob cannot confirm $\beta$ has spin values in both $z$ and $x$ directions simultaneously. The reality that $\beta$ simultaneously have definite values for $\sigma_z$ and $\sigma_x$ cannot be verified. This is consistent with the Heisenberg Uncertainty Principle. There is no incompleteness issue for quantum mechanics. Hence the original EPR argument no longer holds with the modification on the definition of physical realism.

\subsection{Non-causal Correlation}
However, there is still a puzzle here. It appears Bob's measurement outcome on $\beta$ ``depends" on which direction Alice chooses to measure $\alpha$. Since Alice's measurement does not impact the physical property of particle $\beta$, exactly what spin state $\beta$ is in before Alice's measurement? To answer this subtle question, we first note that it is Alice's new gained information on $\beta$, not the physical reality of $\beta$, that depends on the axis along which the measurement is performed. To confirm the new-found reality of $\beta$ relative to Alice, Alice sends the measurement result to Bob who performs a subsequent measurement. There is no faster-than-light action here. One cannot assume there exists an absolute reality for $\beta$. Secondly, it is true that Bob's measurement outcome correlates to the Alice's measurement result. But this is an informational correlation, not a causal relation. This correlation is encoded in the entangled state of the composite system $\alpha + \beta$ described in (\ref{EPR}). Since the entanglement is preserved even when both particles are space-like separated, the correlation is preserved. Such entangled quantum state encode not only the classical correlation, but also the additional information of the composite system (see explanation in Appendix F). When Alice measures particle $\alpha$, she effectively measures the composite system, because she obtains information not only about $\alpha$, but also about the correlation between $\alpha$ and  $\beta$. In addition, the measurement induces decoherence of the $\alpha + \beta$ composite system. Before Alice performs the measurement, it is meaningless to speculate what spin state particle $\beta$ is in since it is in a maximum mixed state. When Alice measures $\alpha$ along $z$ direction and obtains result of spin up, she knows that in this condition, $\beta$ is also in spin up and later this is confirmed by Bob. If instead, she measures $\alpha$ along $x$ direction and obtains result of spin left, she knows that in this new condition, $\beta$ is in spin left and later confirmed by Bob. But such correlation is not a causal relation. To better understand this non-causal relation, supposed there are $N$ identical but distinguishable copies of the entangled pairs described by Eq. (\ref{EPR}) and each pair has a label $n=1,2,3,\ldots, N$. Alice measures the $\alpha$ particles sequentially along the $z$ direction and she does not send measurement results to Bob. Bob independently measures the $\beta$ particles along $z$ direction with same sequence. Both of them observe their own measurement results for $\sigma_z$ as randomly spin up or spin down with fifty percent of chance for each. When later they meet and compare measurement results, they find the sequence of $\sigma_z$ values are exactly the matched. They can even choose a random sequence of $z$ or $x$ direction but both follow the exact sequence in their independent measurements. When later they meet and compare measurement results, they still find their measured values are the same sequentially. 

\subsection{Implications}
The EPR experiment shows that a quantum measurement should be explicitly described as observer dependent. The idea of observer-independent quantum state should be abandoned since it assumes there exists a Super Observer. Assumption of a Super Observer is non-operational since practically a physical observer is always local. When we measure a microscopic object which is much smaller than the apparatus, the apparatus can detect change in any part of the system under observation. This is what quantum mechanics was originally developed for. However, when one wishes to extend the quantum description to a system that is spatially much larger than a typical apparatus system in a measurement, the locality of the apparatus becomes important. The implication here is that information exchange between quantum systems is local. Description of a quantum process involving information exchange between quantum systems must factor in such locality principle.

\section{Discussion and Conclusion}
\label{sec:conclusion}

\subsection{Conceptual Consequences}
The relational formulation of quantum measurement results in many conceptual implications. We will highlight the key conceptual consequences in this section. 

\textit{Explicit relativity of measurement process.}
A key conclusion of this work is that for a given quantum system, it time evolution process is implicitly relative, while a measurement process is explicitly relative. The criterion to distinguish time evolution and measurement is whether the entanglement entropy of the observed system changes. Therefore, if the entanglement entropy changes, the process is explicitly relative. 

\textit{Relativity of Information.} The notion of information here refers to the correlation between the observed system and the measuring system, and is measured by the entropy of reduced density matrix of the observed system. Change of the entropy means change of information. Thus, a quantum process to extract information from a system must be described explicitly relative to an observer. Consequently, there is no absolute information to all observers in quantum mechanics, just as there is no absolute spacetime in Relativity. Even when the time evolution is described by the Schr\"{o}dinger Equation without calling out the observer explicitly, the underlined notion that the quantum theory is relative should not be forgotten. 

\textit{No Super Observer.}
The assumption that there exists a Super Observer can revert explicit relativity back to implicit relativity, because a Super Observer instantaneously knows the outcome of a measurement performed at any remote location. One can always describe a process referring to the Super Observer. Thus, there is no need to call out observer explicitly which in turns allows absolute quantum state. However, the assumption of Super Observer is non-operational, because practically a physical apparatus is practically finite and local. Consequently, the measurement event is also local.

\textit{Objectivity of a quantum state.} The relational nature of a quantum state does not imply a quantum state is subjective. Space-like separated local observers can reconcile the different descriptions of the same quantum system through classical communication of information obtained from local measurements, as shown in the analysis of EPR paradox. This is significant since it gives the meaning of objectivity of a quantum state. Objectivity can be defined as the ability of different observers coming to a consensus independently~\cite{Zurek03}. This synchronization of latest information is operational, and it is necessary for consistent descriptions of the same quantum system from different observers. Exactly what the mechanism is to achieve synchronization is not specified here, but such a mechanism must follow the quantum theory.

\subsection{Predictability}
One critical question to ask is in what situation the RQM formulation can predict results different from traditional quantum mechanics. Is it yet another quantum mechanics interpretation only, or is there new physics underlined the formulation? To answer the question, we need to know the limitation of traditional quantum mechanics that RQM helps to remove. Quantum mechanics was initially developed as a physical theory to explain results of observation of microscopic systems, such as spectrum of light emitted from hydrogen atoms. In such condition, the observed system as a whole is much smaller than the apparatus. An observer can read the results at once even when the observed system consists multiple subsystems. The assumption of Super Observer becomes operational, even though it is conceptually incorrect. The assumption of Super Observer practically makes explicit relativity unnecessary. Thus, there is no need to rely on a relational formulation of quantum mechanics. Traditional quantum mechanics is adequate to describe the microscopic physical world in this condition. 

However, when one wishes to construct a quantum theory for composite system that is spatially larger than the typical measuring apparatus by orders of magnitude, REQ formulation becomes necessary, as manifested in the EPR analysis. A typical procedure to construct a quantum description is to define the boundary of the system such that it can be approximated as an isolated system\footnote{Such an approximation may not be always possible if the interaction from the environment cannot be isolated. Suppose the observed system $S$ is interacting with the environmental system $E$. Changing the boundary to include $E$, we have $S+E$. But $S+E$ is interacting with a larger surrounding environmental system $E'$, and so on. One cannot describe the composite system as unitary process unless extending the boundary to the whole Universe.}. Then, for a given set of information, such as an initial quantum state and the Hamilton operator, its time evolution is described as a unitary process. If, however, an event occurs such that one of the subsystem starts to interact with another system outside the composite system and causes information exchange, it must be described as explicitly relative. A different observer who does not know the event, must exchange information with observer who knows the event through classical communication or additional measurement if we expect these two observers to have equivalent descriptions of the same quantum process. The need of information synchronization, which is a result of our relational formulation, becomes a necessary component for an accurate quantum description. Global quantum operation on such a composite system is not practical. If indeed there is a need to describe results from a global operation, extra caution must be taken care to synchronize the measurement outcomes from different remote subsystems.

Having equivalent description of a physical law from different observers is a basic requirement in the Relativity Theory. How quantum measurement is described in the context of Relativity? This is an interesting question to investigate, given that a quantum measurement must be described as observer dependent. We speculate that the need for information synchronization in a quantum measurement is a necessary step when one wishes to combine quantum mechanics with the Relativity Theory.

\subsection{Comparison with the Original RQM Theory}
The works presented here is inspired by the main idea of the original RQM theory~\cite{Rovelli96}. However, there are several significant improvements that should be pointed out. 

The works of Refs.~\cite{Rovelli96, Zurek82, Transs2018} establish the idea that relational properties are more basic, and a quantum system must be described relative to another quantum system. However, they do not provide a clear formulation on how a quantum system should be described relative to another system and what the basic relational properties are. On the other hand, our formulation gives a clear quantification of the relational property, which is the relational probability amplitude. The introduction of the relational probability amplitude is based on a detailed analysis of measurement process. It enables us to develop a framework to calculate probability during quantum measurement. We further show that the relational probability amplitude can be calculated using Feynman path integral~\cite{Yang2021}. 

The second improvement in this works comes from the introduction of the concept of entanglement to the RQM theory. We recognize not only that a quantum system must be described relative to another quantum system, but also that the entanglement between these two systems impacts the formulation the observed system is described. If there is no entanglement, the observed system can be described by a wave function. If there is entanglement, a reduced density matrix is more appropriate mathematical tool. In addition, entanglement measure plays a pivot role in determining a system is undergoing a time evolution or measurement process. This allows us to reconstruct both the Schr\"{o}dinger equation~\cite{Yang2017} and the measuring theory. When one states that a quantum system must be described relative to another quantum system, one can further quantify this relativity via the entanglement measure between these two systems. However, the concept of entanglement is not presented in Ref~\cite{Rovelli96}. The reconstruction attempts in Ref~\cite{Rovelli96, Transs2018} to derive the laws of quantum mechanics based on quantum logic does not yet include quantum measurement theory. 

Thirdly, although a quantum system must be described relative to another quantum system, our work shows that there are mathematical tools that can describe the observed system without explicitly calling out the reference system, such as the wave function or the reduced density matrix. Therefore, RQM and traditional QM are compatible mathematically. We further show that for a given system, its time evolution process is implicitly relative, while a measurement process is explicitly relative. The measurement process, when applied to microscopic physical system that are much smaller than the apparatus, can be practically described without calling out the reference system. In such cases, RQM and traditional QM are practically equivalent. This is important because it confirms that although the main idea of RQM seems radical, it does not change the practical application of quantum mechanics. These points were not clear in Ref~\cite{Rovelli96}. 

\subsection{Relation to QRF Theory}
As discussed in the introduction section, the RQM principle consists two aspects. First, we need to reformulate quantum mechanics relative to a QRF which can be in a superposition quantum state, and show how quantum theory is transformed when switching QRFs. In this aspect, we accept the basic quantum theory as it is, including Schr\"{o}dinger equation, Born's rule, von Neunman measurement theory, but add the QRF into the formulations and derive the transformation theory when switching QRFs~\cite{Brukner, Hoehn2018, Yang2020}. Second, we go deeper to reformulate the basic theory itself from \textit{relational properties}, but relative to a fixed QRF. Here the fixed QRF is assume to be in a simple eigen state. This is what we do in Ref.~\cite{Yang2017} and the present work. A complete RQM theory should combine these two aspects together. This means one will need to reformulate the basic quantum theory from relational properties and relative to a quantum reference frame that exhibits superposition behavior. Therefore, a future step is to investigate how the relational probability amplitude matrix should be formulated when the QRF possesses superposition properties, and how the relational probability amplitude matrix and the measurement formulations are transformed when switching QRFs.

\subsection{Conclusions}
Quantum measurement and quantum operation theory is developed here based on the relational formulations of quantum mechanics~\cite{Yang2017}. The relational properties are the starting point to construct the quantum measurement and quantum operation theory. We show that how the relational probability amplitude matrix is transformed and how mutual information is exchanged during measurement. The resulting formulation is mathematically compatible with the traditional quantum mechanics.   

The significance of our formulation comes from the conceptual consequences. We assert that for a given quantum system, description of its time evolution can be implicitly relative, while description of a quantum operation must be explicitly relative. Information exchange is relative to a local observer in quantum mechanics. The assumption of Super Observer should be abandoned, so as the notion of observer independent description of physical reality. Different local observers can achieve consistent descriptions of a quantum system if they are synchronized on the outcomes from any measurement performed on the system, thus achieve an objective description. The conceptual subtlety of the relativity and objectivity of a quantum description is not obvious to recognize in traditional quantum mechanics, because traditional quantum mechanics was originally developed to explain observation results from microscopic system that is much smaller than the measuring apparatus. For those situations, RQM and traditional quantum mechanics are practically equivalent. However, for a composite system that is spatially much larger than a typical apparatus, the necessity of RQM formulation becomes clear, as manifested in the analysis of EPR paradox. The paradox is seemingly inevitable in traditional quantum mechanics but can be resolved by removing the assumption of the Super Observer. The completeness of quantum mechanics and locality can coexist by redefining the element of physical reality to be observer-dependent. There might be more results from this direction. We further speculate that the synchronization of measurement results from different observers is a necessary step when combining quantum mechanics with relativity theory.

As stated philosophically in Ref.~\cite{Smolin}, the physical world is made of processes instead of objects, and the properties are described in terms of relationships between events. Based on the initial RQM reformulation effort~\cite{Rovelli96}, Ref.~\cite{Yang2017} and this paper together further show that quantum mechanics can be constructed by shifting the starting point from the independent properties of a quantum system to the relational properties among quantum systems. The reformulation results in more clarity of many subtle physical concepts. We envision next step is to investigate how the relational probability amplitude matrix and the measurement formulations are transformed when switching QRFs. It is our belief that these efforts together is one step towards a better understanding of quantum mechanics.



%
%




\appendix
\section{Theorem 1}
\label{Theorem1}
\begin{theorem} 
$H(R)=0$ if and only if the matrix element $R_{ij}$ can be decomposed as $R_{ij}=c_id_j$, where $c_i$ and $d_j$ are complex numbers.
\end{theorem}
\textbf{Proof:} According to the singular value decomposition, the relational matrix $R$ can be decomposed to $R = UDV$, where $D$ is rectangular diagonal and both $U$ and $V$ are $N\times N$ and $M\times M$ unitary matrix, respectively. This gives $\rho = RR^\dag = U(DD^\dag) U^\dag$. If $H(R)=0$, matrix $\rho$ is a rank one matrix, therefore $DD^\dag$ is $diag\{1,0,0...\}$. This means $D$ is a rectangular diagonal matrix with only one eigenvalue $e^{i\phi}$. Expanding the matrix product $R=UDV$ gives 
\begin{equation}
    R_{ij}=\sum_{nm}U_{in}D_{nm}V_{mj}=U_{i1}e^{i\phi}V_{1j}.
\end{equation}
We just choose $c_i=U_{i1}$ and $d_j=e^{i\phi}V_{1j}$ to get $R_{ij}=c_i d_j$. Conversely, if $R_{ij}=c_i d_j$, $R$ can be written as outer product of two vectors,
\begin{equation}
R = \begin{pmatrix} c_1 & c_2 & \ldots  & c_n \end{pmatrix}^T \times \begin{pmatrix} d_1 & d_2 & \ldots & d_m \end{pmatrix}.
\end{equation}
Considering vector $C_1=\{c_1, c_2, \ldots,c_n\}$ as an eigenvector in Hilbert space ${\cal H}_S$, one can use the Gram-Schmidt procedure~\cite{Nelson00} to find orthogonal basis set $C_2, \ldots, C_n$. Similarly, considering vector $D_1=\{d_1, d_2, \ldots,d_m\}$ as an eigenvector in Hilbert space ${\cal H}_A$, one can find orthogonal basis set $D_2, \ldots, D_m$. Under the new orthogonal eigenbasis, $R$ becomes a rectangular diagonal matrix $D=diag\{1,0,0...\}$. Therefore $R=UDV$ where $U$ and $V$ are two unitary matrices associated with the eigenbasis transformations. Then $\rho = RR^\dag=U(DD^\dag) U^\dag$, and $DD^\dag=diag\{1,0,0...\}$ is a square diagonal matrix. Since the eigenvalues of similar matrices are the same, the eigenvalues of $\rho$ are (1, 0, ...), thus $H(R)=0$. 

\section{Decomposition of the Unitary Operator of a Bipartite System}
\label{Decomposition}
If there is interaction between $S$ and $A$, and the initial state of $S+A$ is a product state, the global unitary operator can be decomposed into a set of measurement operators that satisfies Eq. (\ref{map1}). The proof shown here closely follows idea from Ref.~\cite{Nielsen001}. Denote the initial state is product state, $|\Psi_0\rangle=|\psi_0\rangle_S|\phi_0\rangle_A$. First we change the eigenbasis for $A$ through a local unitary operator $I_S\otimes \hat{U}_A$ such that $\phi_0$ is the first eigenvector of the orthogonal eigenbasis, i.e., $(I_S\otimes \hat{U}_A)|\psi_0\rangle_S \otimes|\phi_0\rangle_A = |\psi_0\rangle_S\otimes|a_0\rangle_A$, and $\{|a_m\rangle \}$ forms an orthogonal eigenbasis of $A$. The global unitary operator is changed to $\hat{U}_{SA}^{'} = \hat{U}_{SA}(I_S\otimes \hat{U}_A^\dag)$. Define a linear operator $\hat{M}_m = \langle a_m|\hat{U}_{SA}^{'}| a_0\rangle$. The set of operators $\{\hat{M}_m\}$ is what we are looking for, since we can verify it satisfies Eq. (\ref{map1}),
\begin{equation}
\begin{split}
 \hat{U}_{SA}|\psi_0\rangle|\phi_0\rangle & =\hat{U}_{SA}^{'}|\psi_0\rangle|a_0\rangle \\
 & = \sum_m |a_m\rangle\langle a_m| \hat{U}_{SA}^{'}|\psi_0\rangle|a_0\rangle \\
 & = \sum_m\hat{M}_m|\psi_0\rangle|a_m\rangle.
\end{split}
\end{equation}
The completeness condition can also be verified,
\begin{equation}
\begin{split}
    \sum_m\hat{M}_m^\dag\hat{M}_m & = \sum_m\langle a_0|\hat{U}_{SA}^{'\dag}|a_m\rangle\langle a_m|\hat{U}_{SA}^{'}|a_0\rangle \\
    & = \langle a_0|\hat{U}_{SA}^{'\dag}\hat{U}_{SA}^{'}|a_0\rangle = I_S.
\end{split}
\end{equation} 

\section{Theorem 2}
\label{Theorem2}
\begin{theorem}
Applying operator $\hat{Q}\otimes \hat{O}$ over the composite system $S+A$ is equivalent to change the relational matrix $R$ to $R'=QRO^T$, where the superscript $T$ represents a transposition.
\end{theorem}
\textbf{Proof:} Denote the initial state vector of the composite system as $|\Psi_0\rangle=\sum_{ij}R_{ij}|s_i\rangle|a_j\rangle$. Apply the composite operator $\hat{Q}(t)\otimes\hat{O}(t)$ to the initial state,
\begin{equation}
\begin{split}
 |\Psi_1\rangle & = (\hat{Q}\otimes \hat{O}) \sum_{ij} R_{ij} |s_i\rangle \otimes |a_j\rangle \\
 & = \sum_{ij}R_{ij}\hat{Q}|s_i\rangle\otimes \hat{O}|a_j\rangle \\
 & = \sum_{ij}\sum_{mn}R_{ij}Q_{mi}O_{nj}|s_m\rangle\otimes |a_n\rangle \\
 & = \sum_{mn}(\sum_{ij}Q_{mi}R_{ij}O^T_{jn})|s_m\rangle\otimes |a_n\rangle \\
 & =\sum_{mn}(QRO^T)_{mn}|s_m\rangle|a_n\rangle
\end{split}
\end{equation}
where $T$ represents the transposition of matrix. Compared the above equation to Eq.(\ref{WF}) for the definition of $|\Psi_1\rangle$, it is clear that the relational matrix is changed to $R'=QRO^T$.

\section{Probability in Selective Measurement}
Given the composite system $S+A$ is described by Eq.(\ref{WFSA}),
the reduced density matrix of $S$ can be defined
\begin{equation}
    \begin{split}
    \hat{\rho}_S & =Tr_A|\Psi\rangle\langle\Psi| = \sum_{ii'}(\sum_kR_{ik}R^*_{i'k})|s_i\rangle\langle s_{i'}| \\
    & =\sum_{ii'}(RR^\dag)_{ii'}|s_i\rangle\langle s_{i'}|
    \end{split}
\end{equation}
and the probability of finding event $|s_i\rangle$ occurred to $S$ is calculated by Eq.(\ref{indirectProb}). Similarly, the probability of event $|a_j\rangle$ occurred to $A$ is $p^A_j=\sum_ip_{ij}=\sum_i|R_{ij}|^2$. This can be more elegantly written by introducing a partial projection operator $I^S\otimes \hat{P}_j^A$ where $\hat{P}_j^A=|a_j\rangle\langle a_j|$. It is easy to verify that
\begin{equation}
\label{partialProb}
\begin{split}
p_j^A & =\langle\Psi|I^S\otimes \hat{P}_j^A|\Psi\rangle \\
& =\langle\Psi|a_j\rangle\langle a_j|\Psi\rangle =\sum_i|R_{ij}|^2.
\end{split}
\end{equation}

\section{Open Quantum System}
\label{sec:OQS}
The measurement theory described in this paper is consistent with the open quantum system (OQS) theory. OQS studies the dynamics when a quantum system interacts with its environment $E$~\cite{Nelson00, Rivas12, Breuer07}. Such interaction can result in entanglement and information exchange between the quantum system $S$ and the environment system $E$. Recall the definition of apparatus in Section \ref{subsec:definition} includes the interacting environment as one type of apparatus, if we replace the environment system $E$ with the apparatus system $A$, the OQS theory gives the same formulations as shown in Section \ref{sec:measurement}.

First, we give a brief review of the OQS theory. Suppose the initial composite state for the quantum system and environment is described by a density matrix $\rho_{SE}$, the interaction between $S$ and $E$ changes the density matrix $\rho_{SE} \rightarrow \hat{U}\rho_{SE}\hat{U}^\dag$. The resulting density matrix of $S$ is $\rho^{\prime}_S = Tr_E(\hat{U}\rho_{SE}\hat{U}^\dag)$. 
Denote the orthogonal eigen basis of the environment $\{|e_k\rangle\}$. Since the orthogonal eigenbasis of the environment is not necessary the same eigenbasis that diagonalizes $\rho_E$, we further denote the spectral decomposition of $\rho_E$ as $\rho_E =\sum_m\lambda_m|\tilde{e}_m\rangle\langle \tilde{e}_m|$. Assumed the initial state of the quantum system plus environment is a product state, i.e., $\rho_{SE}=\rho_S\otimes \rho_E$, the density operator of $S$ after the interaction with the environment is
\begin{equation}
    \label{OpenQS2}
    \begin{split}
    \rho^{\prime}_S & = \Lambda (\rho_S) \\
    & = \sum_{mk}|\lambda_m|^2 \langle e_k| \hat{U}(\rho_S\otimes |\tilde{e}_m\rangle\langle \tilde{e}_m| )\hat{U}^\dag|e_k\rangle \\
    & = \sum_{mk}E_{mk}\rho_S E_{mk}^\dag.
    \end{split}
\end{equation}
where $E_{mk}=\lambda_m\langle e_k| \hat{U}|\tilde{e}_m\rangle$ and satisfies the completeness condition $\sum_{mk}E_{mk}E_{mk}^\dag =I$. Eq.(\ref{OpenQS2}) is the Kraus representation of the linear map $\Lambda$. It is proved that $\Lambda$ can be a Kraus representation if and only if it can be induced from an extended system with initial condition $\rho_{SE}=\rho_S\otimes \rho_E$~\cite{Rivas12}. If $\rho_E=|e_0\rangle\langle e_0|$ is a pure state, the linear map is further simplified to $\Lambda (\rho_S)=\sum_kE_{k}\rho_S E_{k}^\dag$ and $E_k=\langle e_k| \hat{U}|e_0\rangle$. The operator set $\{E_k\}$ forms a POVM, and $\Lambda (\rho_S)$ is a Complete Positive Trace Preserving (CPTP) map~\cite{Hayashi15}. To connect to the measurement theory, suppose the measurement outcome $m$ corresponds to an orthogonal state $|\phi_m\rangle$ of $E$, and represented by a projection operator $\hat{P}_m=|\phi_m\rangle\langle\phi_m|$, 
\begin{equation}
    \label{OpenQS3}
    \begin{split}
\rho_S^m & =\Lambda_m (\rho_S) \\
& = \sum_k\langle e_k| \hat{P}_m\hat{U}\rho_S\otimes |\tilde{e}_0\rangle\langle \tilde{e}_0| \hat{U}^\dag\hat{P}_m^\dag|e_k\rangle \\
& = \langle\phi_m|\hat{U}|\tilde{e}_0\rangle\rho_S\langle \tilde{e}_0| \hat{U}^\dag|\phi_m\rangle\sum_k\langle k|\phi_m\rangle\langle\phi_m|k\rangle \\
& = \hat{M}_m\rho_S \hat{M}_m^\dag
    \end{split}
\end{equation}
where $\hat{M}_m=\langle\phi_m|\hat{U}|\tilde{e}_0\rangle$ is the operator defined on $\cal{H}_S$. The probability of finding outcome $m$ is $p_m=Tr(\Lambda_m (\rho_S))=Tr(\hat{M}_m \hat{M}_m^\dag\rho_S)$.

It is evident that if we replace the environment system $E$ with the apparatus system $A$, the OQS theory gives the same formulations as shown in Section \ref{sec:measurement}. In the case of initial product state, Eqs. (\ref{measureWF1}) versus (\ref{OpenQS3}) are effectively the same. Let's consider the case of initial entangled state in the OQS context. Denote the initial system plus environment state as pure bipartite state $|\Psi_0\rangle$. After the global unitary operation $\hat{U}$ and subsequent projection $I^S\otimes \hat{P}_m^E=I^S\otimes|\phi_m\rangle\langle\phi_m|$, the composite state becomes $|\Psi_1\rangle=(I^S\otimes \hat{P_m^E})\hat{U}|\Psi_0\rangle$, take the partial trace over $E$, we get
\begin{equation}
    \label{OpenQSMap}
    \begin{split}
    \rho^S_m  & =Tr_E(|\Psi_1\rangle\langle\Psi_1| \\
    & = \langle\phi_m|\hat{U}_{SA}|\Psi_0\rangle\langle\Psi_0|\hat{U}_{SA}^\dag|\phi_m\rangle \\
    & = |\psi_m\rangle\langle\psi_m|.
    \end{split}
\end{equation}
This implies the resulting state for $S$ is $|\psi_m\rangle=\langle\phi_m|\hat{U}_{SA}|\Psi_0\rangle$, which is equivalent to Eq.(\ref{measureWF3}). Taking a similar approach in deriving Eq.(\ref{measureWF4}), we can express $|\psi_m\rangle$ in the eigenbasis derived from the Schmidt decomposition of $|\Psi_0\rangle$. The result is $|\psi_m\rangle = 1/\sqrt{p'_m} \sum_i \hat{M}_{mi}|\tilde{s}_i\rangle$, where the definitions of $\hat{M}_{mi}$ and $p'_m$ are the same as those in Eq.(\ref{measureWF4}) except replacing the apparatus system $A$ with the environment system $E$. The reduced density operator for $S$, $\rho^S_m$, is given by
\begin{equation}
    \label{OpenQSMap2}
    \begin{split}
    \rho_S^m  = \frac{1}{p'_m}\sum_{ij}\hat{M}_{mi}|\tilde{s}_i\rangle\langle \tilde{s}_j|\hat{M}_{mj}^\dag 
    \end{split}
\end{equation} 
Eq. (\ref{OpenQSMap2}) can be considered as a generalization of Eq.(\ref{OpenQS3}) when the initial state is entangled. There is no simple form of map $\Lambda_m$ such that $\rho_S^m=\Lambda_m(\rho_S)$ where $\rho_S$ is the initial density matrix of $S$. A different representation of $\rho_S^m$ can be derived by rewriting the initial state $\rho_{SE}=\rho_S\otimes\rho_E + \rho_{corr}$ where $\rho_{corr}$ is a correlation term~\cite{Rivas12}. 

\section{Mutual Information}
Entanglement between the two systems is measured by the parameter $E(\rho)$ as defined by Eq.(\ref{vonNeumann}). In Section \ref{sec:measurement}, we also use the mutual information variable to measure the information exchange between the measuring system and the measured system. The mutual information between $S$ and $A$ is defined as $I = H(\rho_S) + H(\rho_A) - H(\rho_{SA})$, where $H(\rho_{SA})$ is the von Neumann entropy of the composite system $S+A$. For a composite system $S+A$ that is described by a single relational matrix $R$, these two variables differ only by a factor of two. However, for a composite system of $S+A$ that is described by an ensemble of relational matrices, the two variables can be very different. This is illustrated by two examples described below.

\textit{Case 1.} $S+A$ is in an entangled pure state described by $|\Psi\rangle_{SA}=\sum_i\lambda_i|s_i\rangle|a_i\rangle$ in Schmidt decomposition, where $\{\lambda_i\}$ are the Schmidt coefficients. Subsystem $S$ is in a mixed state. The entanglement measure is $E(\rho_S)= -\sum_i\lambda_i^2ln(\lambda_i^2)$ and the mutual information is $I=-2\sum_i\lambda_i^2ln(\lambda_i^2)$. 

\textit{Case 2.} $S+A$ is in a mixed state described by $\rho_{SA}=\sum_i\lambda_i^2|s_i\rangle|a_i\rangle\langle s_i|\langle a_i|$. In the case, $H(\rho_S)=H(\rho_{SA})=H(\rho_{diag})=-\sum_i\lambda_i^2ln(\lambda_i^2)$. $\rho_{SA}$ is a separable bipartite state~\cite{Hayashi15, Horodecki}. There is no entanglement but there is mutual information since $I=-\sum_i\lambda_i^2ln(\lambda_i^2)$. Essentially the composite system is a mixed ensemble of product states $\{\lambda_i^2, |s_i\rangle|a_i\rangle\}$. One can infer that $S$ is in $|s_i\rangle$ from knowing $A$ is in $|a_i\rangle$, however such mutual information is due to classical correlation. The probability of finding $S$ in an eigenvector $|s_i\rangle$ is just the classical probability $\lambda_i^2$. 

Although the reduced density operator for $S$, $\rho_S = \sum_i\lambda_i^2|s_i\rangle\langle s_i|$, is the same in \textit{Case 1} and \textit{Case 2}, the mutual information is different. More information is encoded in the pure bipartite state in \textit{Case 1}. When $S+A$ is described by $|\Psi\rangle_{SA}=\sum_i\lambda_i|s_i\rangle|a_i\rangle$, besides the inference information between $S$ and $A$, there is additional indeterminacy due to the superposition at the composite system level. For instance, one cannot determine the composite system $S+A$ is in $|s_0\rangle|a_0\rangle$ or $|s_1\rangle|a_1\rangle$ before measurement. More indeterminacy before measurement means more information can be gained from measurement. This also explains that in the EPR experiment, when Alice measures particle $\alpha$, she does not only gain information about $\alpha$, but also gain information about the composite system. Thus, she can predict the state of particle $\beta$. On the other hand, such indeterminacy does not exist when $S+A$ is described by a mixture of product state as in \textit{Case 2}. Since such indeterminacy is for the composite system as a whole, the reduced density operator for a subsystem $S$, $\rho_S$, cannot reflect the difference, therefore it appears the same in \textit{Case 1} and \textit{Case 2}.  

These two examples show that mutual information variable can substitute the entanglement measurement only when the composite system $S+A$ is described by a single relational matrix $R$. To quantify change of quantum correlation during a measurement, the entanglement measurement is a more appropriate parameter. 

\end{document}